\documentclass[preprint,review,11pt]{elsarticle}

\journal{Reliability Engineering \& System Safety}

\usepackage{fullpage}
\usepackage{color,soul}
\usepackage{ulem}
\usepackage{rotating}
\usepackage{graphicx,psfrag,epsf}
\usepackage{enumerate}
\usepackage{float}
\usepackage{lineno,xcolor}

\usepackage{amsmath}
\usepackage{amssymb}

\usepackage{url} 
\usepackage{subfigure}

\newcommand{\blind}{0}

\begin{document}


\if0\blind
{
  \title{The generalization of Latin hypercube sampling}
  \author{Michael D.\ Shields$^{*}$ and Jiaxin Zhang  \\
    Department of Civil Engineering, Johns Hopkins University\\
    $~^{*}$ Corresponding Author: michael.shields@jhu.edu \\}
  
} \fi

\if1\blind
{
  \bigskip
  \bigskip
  \bigskip
  \begin{center}
    {\LARGE\bf The generalization of Latin hypercube sampling}
\end{center}
  \medskip
} \fi

\bigskip


\begin{abstract}
Latin hypercube sampling (LHS) is generalized in terms of a spectrum of stratified sampling (SS) designs referred to as partially stratified sample (PSS) designs. True SS and LHS are shown to represent the extremes of the PSS spectrum. The variance of PSS estimates is derived along with some asymptotic properties. PSS designs are shown to reduce variance associated with variable interactions, whereas LHS reduces variance associated with main effects. Challenges associated with the use of PSS designs and their limitations are discussed. To overcome these challenges, the PSS method is coupled with a new method called Latinized stratified sampling (LSS) that produces sample sets that are simultaneously SS and LHS. The LSS method is equivalent to an Orthogonal Array based LHS under certain conditions but is easier to obtain. Utilizing an LSS on the subspaces of a PSS provides a sampling strategy that reduces variance associated with both main effects and variable interactions and can be designed specially to minimize variance for a given problem. Several high-dimensional numerical examples highlight the strengths and limitations of the method. The Latinized partially stratified sampling method is then applied to identify the best sample strategy for uncertainty quantification on a plate buckling problem.
\end{abstract}


\maketitle


\section{Introduction}
Latin hypercube sampling (LHS) \cite{McKay_et_al_Tech_79,Helton_Davis_RESS_03} is the most widely used random sampling method for Monte Carlo-based uncertainty quantification, employed in nearly every field of computational science, engineering, and mathematics. The seminal work by McKay et al.\ \cite{McKay_et_al_Tech_79} introducing Latin hypercube sampling is a classic in the field of design of computer experiments. LHS is an especially powerful and useful sampling method thanks primarily to the properties identified by Stein \cite{Stein_Tech_87} who showed that LHS has the effect of filtering the variance associated with the additive components of a transformation (or main effects). This result, combined with the Hierarchical Ordering Principle \cite{Wu_Hamada_00} - which states that main effects and low order interactions are likely to govern most general transformations - causes LHS to reduce variance significantly for many applications.

The widespread popularity of LHS has led to the invention of numerous variants meant to improve space-filling \cite{Johnson_et_al_JSPI_90, Morris_Mitchell_JSPI_95,Ye_JASA_98,Ye_et_al_JSPI_2000,Cioppa_Lucas_Tech_07,Joseph_Hung_SS_08,Dalbey_Karystinos_AIAA_10}, optimize projective properties \cite{Liefvendahl_Stocki_JSPI_06}, minimize least square error and maximize entropy \cite{Park_JSPI_94}, and reduce spurious correlations \cite{Iman_Conover_CSSC_82,Florian_PEM_92,Tang_JASA_93,Huntington_Lyrintzis_PEM_98,Ye_JASA_98,Cioppa_Lucas_Tech_07,Vorechovsky_Novak_PEM_09}. Meanwhile, LHS has been applied to nearly every type of probabilistic analysis one can imagine, ranging from the estimation of reliability (probability of failure) \cite{Olsson_et_al_SS_03} to coefficient estimation for polynomial chaos, neural network, and other types of surrogate models \cite{Ghiocel_Ghanem_JEM_02,Giovanis_Papadopoulos_ES_15}. The intention of this paper is not to present another variant of the LHS methodology or to apply it in a new or novel way. Rather, we present a broad generalization of the methodology in the context of stratified sampling - from which LHS is derived.

\begin{figure}
\centering
\includegraphics[width=1.\columnwidth]{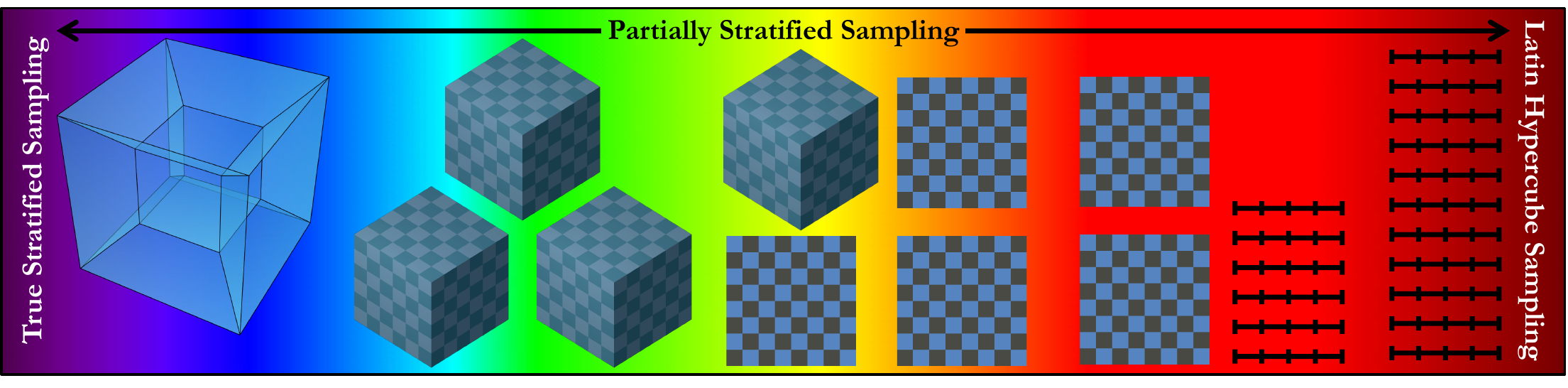}
\caption{Spectrum of stratified sampling methods.}
\label{fig:spectrum}
\end{figure}

Stratified sampling (SS), the ``parent" methodology of LHS has been widely used in the social sciences and financial mathematics owing to its ability to partition a population into strata (or categories) that can be weighted according to their conditional probabilities. Some recent developments have begun to encourage its use in Monte Carlo analysis of computer models by adaptively stratifying the probability space of a random vector \cite{Shields_et_al_RESS_15,Shields_Sundar_IJRS_2015}. These methods rely on ``true" stratified sampling wherein all dimensions of the space are stratified simultaneously allowing the analyst to concentrate samples in probabilistically weighted regions of the space that are important for the problem at hand. LHS, meanwhile, lies at the opposite end of the ``spectrum'' of stratified sampling methods (Figure \ref{fig:spectrum}) such that each dimension of the random vector is stratified individually and the vector is constructed through random pairings. 

In this work, the intermediate space on the spectrum of stratified sampling methods is explored such that stratification can occur on any set of $N_i$-dimensional orthogonal subspaces of the $N$-dimensional sample space $\mathbf{\mathcal{S}}$ subject to $\sum_i N_i=N$. This generalized sampling method is referred to as Partially Stratified Sampling (PSS) and the properties of their designs are explored in detail. In particular, the PSS method is shown to reduce variance associated with low-dimensional interactions within a high-dimensional transformation. Some discussion surrounding the appropriate use of a PSS design is provided and a powerful hybrid PSS-LHS method is proposed that simultaneously reduces variance associated with the main-effects and low-dimensional interactions. This method, referred to as ``Latinized" Partially Stratified Sampling (LPSS), combines the variance reductions of both PSS and LHS to yield a major improvement in sample efficiency. Several high-dimensional demonstration problems are presented and the method is applied to the probabilistic assessment of plate buckling strength where the interaction of geometric and material variables are very important.


\section{Review of Sampling Methods}
This section provides a brief review of the sampling methods used for analysis in this work. We will consider only uncorrelated random variables as it is common practice to maps correlated variables onto a set of uncorrelated ones using, for example, Principal Component Analysis. 

\subsection{Simple Random Sampling}
Classical Monte Carlo methods rely on so-called Simple Random Sampling (SRS) or Monte Carlo Sampling in which realizations of the vector $\mathbf{x}$ (samples) are generated as independent and identically distributed (iid) realizations on $\mathbf{\mathcal{S}}$ with marginal cumulative distribution functions (CDFs) $D_{X_{i}}(\cdot)$ by:
\begin{equation}
\label{eqn:1}
x_i=D_{X_{i}}^{-1}(U_i); i=1,2,\dots,n
\end{equation} 
where $U_i$ are iid uniformly distributed samples on $[0,1]$. The realizations $\mathbf{x}$ are then applied to the system $\mathbf{y}=\mathbf{F}(\mathbf{x})$ and $\mathbf{y}$ is statistically evaluated. 

\subsection{Stratified Sampling}
\label{sec:SS}
Stratified Sampling begins by dividing the sample space $\mathbf{\mathcal{S}}$ into a collection of $M$ disjoint subsets (strata) $\mathbf{\Omega}_k;  k=1,2,\dots,M$ with $\cup_{k=1}^M\mathbf{\Omega}_k=\mathbf{\mathcal{S}}$ and $\mathbf{\Omega}_p\cap\mathbf{\Omega}_q=\emptyset; p\ne q$. Sample realizations from a given stratum $k$, $\mathbf{x}_k=\left\{x_{1k},x_{2k},\dots,x_{Nk}\right\}$, are generated by randomly sampling the vector components according to:
\begin{equation}
\label{eqn:2}
x_{ik}=D_{X_{i}}^{-1}(U_{ik}); i=1,2,\dots,N
\end{equation} 
where $U_{ik}$ are iid uniformly distributed samples on $[\xi_{ik}^l,\xi_{ik}^u]$ with $\xi_{ik}^l=D_{X_{i}}(\zeta_{ik}^l)$ and $\xi_{ik}^u=D_{X_{i}}(\zeta_{ik}^u)$, and $\zeta_{ik}^l$ and $\zeta_{ik}^u$ denote the lower and upper bounds respectively of the $i^{th}$ vector component of stratum $\mathbf{\Omega}_k$. Typically the stratification is performed directly in the probability space meaning that the strata are defined directly by the bounds $\xi_{ik}^l$ and $\xi_{ik}^u$.

\subsection{Latin Hypercube Sampling}
Latin Hypercube Sampling (LHS) operates by dividing the subspace of each vector component $\mathcal{S}_i; i=1,2,\dots,N$ into $M=n$ disjoint subsets (strata) of \emph{equal probability} $\mathbf{\Omega}_{ik};  i=1,2,\dots,N; k=1,2,\dots,M$. Samples of each vector component are drawn from the respective strata according to:
\begin{equation}
\label{eqn:5}
x_{ik}=D_{X_{i}}^{-1}(U_{ik}); i=1,2,\dots,N;  k=1,2,\dots,M
\end{equation} 
where $U_{ik}$ are iid uniformly distributed samples on $[\xi_{k}^l,\xi_{k}^u]$ with $\xi_{k}^l=\dfrac{k-1}{M}$ and $\xi_{k}^u=\dfrac{k}{M}$. The samples $\mathbf{x}$ are assembled by randomly grouping the terms of the generated vector components. That is, a term $x_{ik}$ is randomly selected from each vector component (without replacement) and these terms are grouped to produce a sample. This process is repeated $M=n$ times. 

Because the component samples are randomly paired, an LHS is not unique; there are $(M!)^{N-1}$ possible combinations. With this in mind, improved LHS algorithms iterate to determine optimal pairings according to some specified criteria - such as reduced correlation among the terms or enhanced space-filling properties (e.g.\ \citep{Johnson_et_al_JSPI_90,Florian_PEM_92,Huntington_Lyrintzis_PEM_98,Fang_et_al_MoC_02,Liefvendahl_Stocki_JSPI_06,Cioppa_Lucas_Tech_07,Joseph_Hung_SS_08}).

\subsection{Variance reduction in stratified sampling and Latin hypercube sampling}
Consider the general statistical estimator defined by:
\begin{equation}
T(y_1,\dots,y_n)=\sum_{l=1}^nw_lg(y_l)
\end{equation}
where $y_l=h(x_l)$ and $x_l$ denotes a sample generated using SRS, SS, or LHS, $w_l$ are sample weights, and $g(\cdot)$ is an arbitrary function. Note that when $g(y)=y^r$, $T$ is an estimate of the $r^{th}$ moment and when $g(y)=I\{y\le Y\}$, where $I\{\cdot\}$ denotes the indicator function, $T$ is the empirical CDF. For conventional Monte Carlo analysis, $w_l=\frac{1}{N}\hspace{3pt}\forall l$ and the variance of the statistical estimator ($T_R$) is given by Var$[T_R]=\frac{\sigma^2}{N}$ where $\sigma^2=$Var$[g(Y)]$. Classical Monte Carlo estimates generally serve as the measure by which variance reduction techniques are compared.

SS and LHS are both techniques to reduce the variance of statistical estimators when compared to classical Monte Carlo estimates although they do so through different statistical mechanisms. Stratified sampling has been proven to unconditionally reduce the variance of statistical estimators (denoted $T_S$) when compared to SRS such that the variance reduction depends on the differences between the strata means $\mu_k$ and the overall mean $\tau$ as \cite{McKay_et_al_Tech_79}:
\begin{equation}
\text{Var}[T_S]=\text{Var}[T_R]-\dfrac{1}{n}\sum_{k=1}^Mp_k(\mu_k-\tau)^2
\label{eqn:ss_var}
\end{equation}
when the strata are sampled proportionately such that the number of samples on stratum $k$, $n_k=p_kn$. LHS, on the other hand, reduces variance by creating negative covariance between sample cells such that the variance of a LHS estimator $T_L$ can be expressed as \cite{McKay_et_al_Tech_79}:
\begin{equation}
\text{Var}[T_L]=\text{Var}[T_R]+\dfrac{n-1}{n}\dfrac{1}{n^N(n-1)^N}\sum_R(\mu_i-\tau)(\mu_j-\tau)
\label{eqn:lhs_var}
\end{equation}
where $\tau$ is, again, the mean value, $\mu_i$ is the mean of LHS cell $i$, and $R$ denotes the $n^N(n-1)^N$ pairs $(\mu_i,\mu_j)$ of cells having no cell coordinates in common. Note that the variance is reduced only when the second term (cell covariance) is negative. This, however, is not a restrictive condition as McKay et al.\ \cite{McKay_et_al_Tech_79} have shown that variance is always reduced for monotonic functions $Y=h(\mathbf{X})$ and Stein \cite{Stein_Tech_87} showed further that variance is reduced for any function $h(\cdot)$ having finite second moment when $n>>N$. Moreover, Stein showed that LHS reduces variance by effectively filtering the main effects of the transformation such that the stronger the main effects, the more the variance is reduced. This will be discussed in more detail later.


\section{Partially Stratified Sampling}
True stratified sampling and Latin hypercube sampling can be viewed as two extremes on a spectrum of possible stratification methods (Figure \ref{fig:spectrum}). In true SS, the stratification occurs on all dimensions simultaneously while LHS stratifies each dimension individually. In this regard, the proposed Partially Stratified Sampling (PSS) methodology represents the broader class of all possible stratifications between (and including) these extremes in which stratification occurs on low dimensional orthogonal subspaces of the $N$-dimensional sample space $\mathbf{\mathcal{S}}$. As such, it can be viewed from either perspective as a compromise on true SS or as a generalization of LHS. Using either interpretation, true SS and LHS are special cases of PSS such that stratification occurs on $N$-dimensional and one dimensional spaces respectively. The following presents the PSS methodology and its properties in detail.

\subsection{PSS Method}
Let $\mathbf{\Theta}_i, i=1,\dots,N_s$ denote $N_s$ disjoint $N_i$-dimensional orthogonal subspaces of the $N$-dimensional sample space $\mathbf{\mathcal{S}}$ ($N_i\le N\hspace{3pt}\forall i$) such that $\cup_{i=1}^{N_s}\mathbf{\Theta}_i=\mathbf{\mathcal{S}}$ and $\mathbf{\Theta}_p\cap\mathbf{\Theta}_q=\emptyset; p\ne q$. Note that the general form does not require $N_i=N_j; i\ne j$. PSS divides each subspace $\mathbf{\Theta}_i$ into a collection of $M_i$ disjoint subsets $\mathbf{\Omega}_{ik}; k=1,2,\dots,M_i$. Lower $N_i$-dimensional random samples $\mathbf{x}_{ik}=\{x_{ik1},x_{ik2},\dots,x_{ikN_{i}}\}$ are generated within each stratum $\mathbf{\Omega}_{ik}$ of subspace $\mathbf{\Theta}_i$ according to the stratified sampling method in Section \ref{sec:SS}. Full $N$-dimensional samples $\mathbf{x}$ are assembled by randomly grouping the lower-dimensional samples generated in each subspace. That is, a low-dimensional sample $\mathbf{x}_{ik}$ is randomly selected from each subspace $\mathbf{\Theta}_i$ and these terms are grouped to produce a sample.

\begin{figure}[!ht]
\centering
\includegraphics[width=1.\columnwidth]{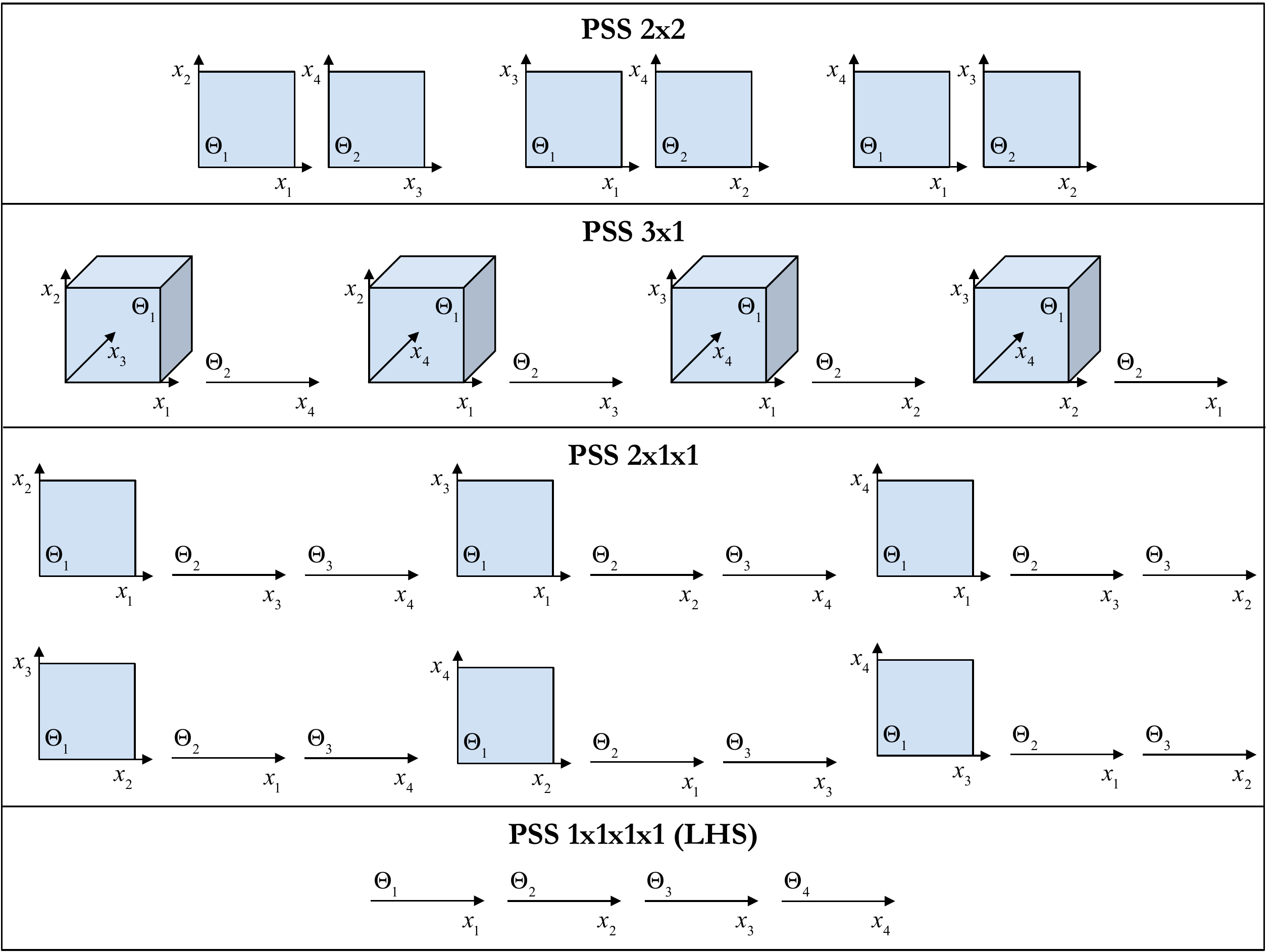}
\caption{Possible PSS subspaces for a 4D sample space.}
\label{fig:4D_PSS}
\end{figure}

\begin{figure}[!ht]
\centering
\includegraphics[width=.6\columnwidth]{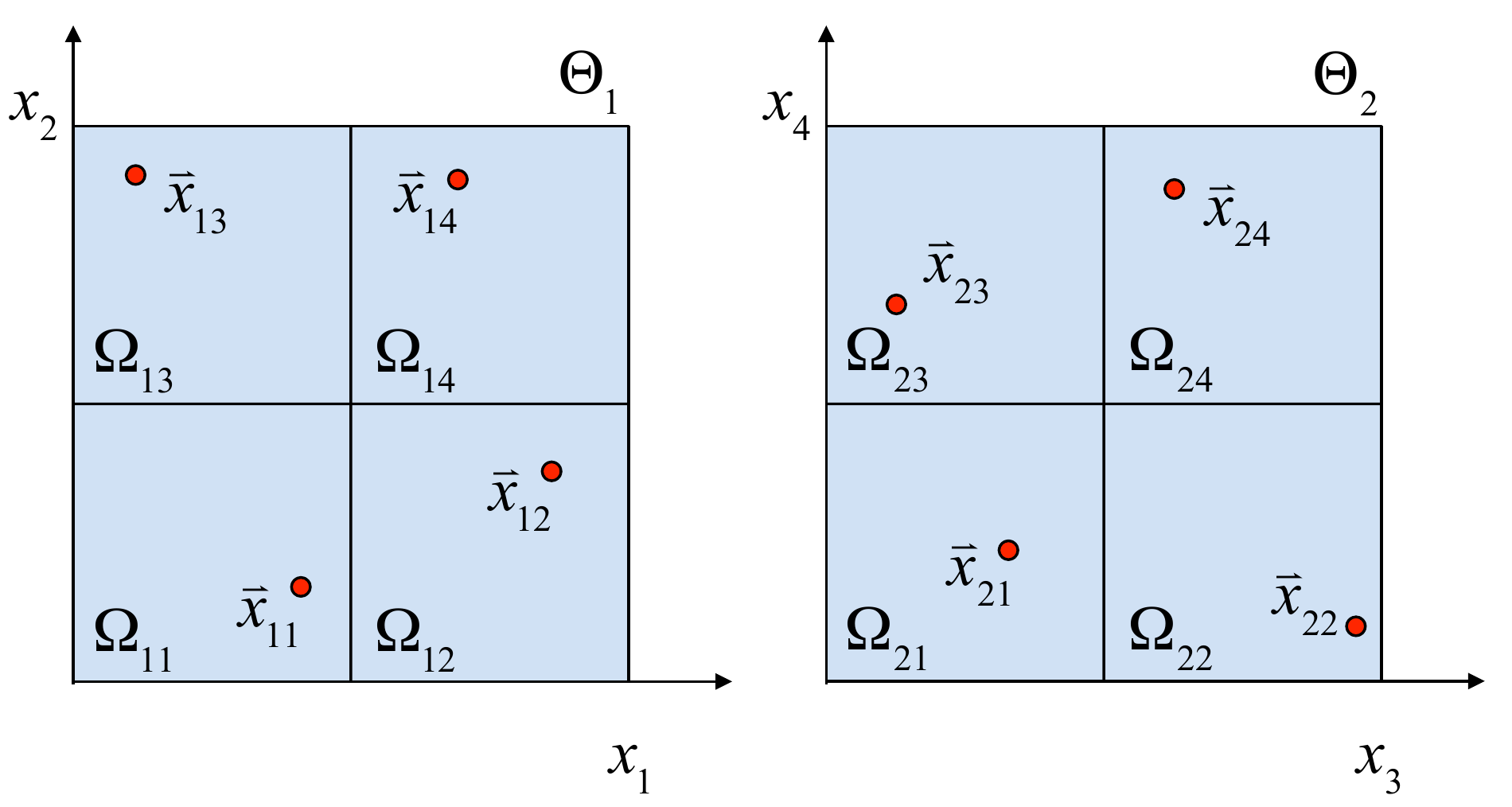}
\includegraphics[width=.6\columnwidth]{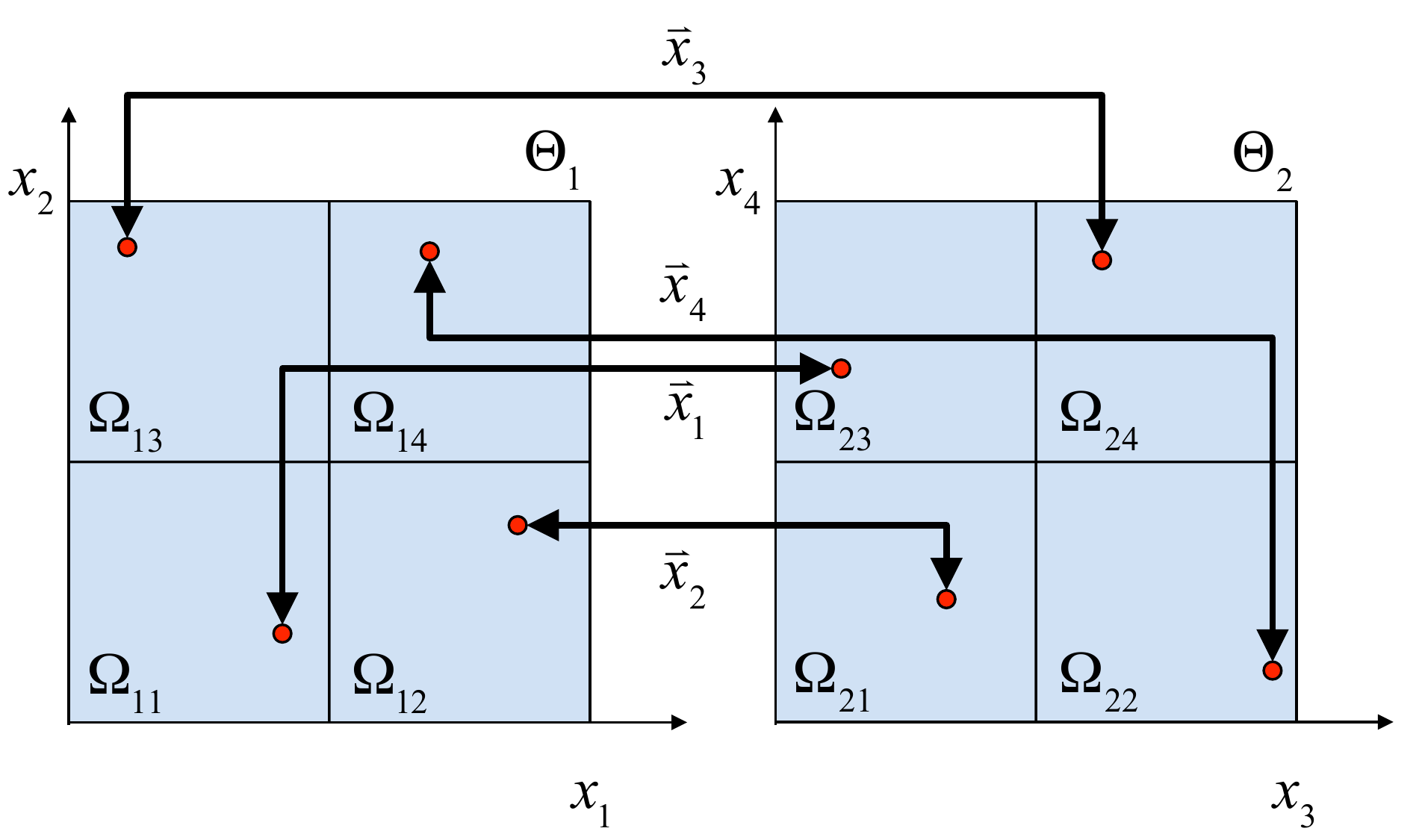}
\caption{Four samples drawn from a 4D sample space using a PSS 2x2. (a.) Draw stratified samples within the 2D subspaces. (b.) Randomly pair the 2D samples to assemble 4D samples.}
\label{fig:2}
\end{figure}

Consider the simple case of a 4-dimensional sample space. Possible PSS subspace designs are shown in Figure \ref{fig:4D_PSS}. 
Next, consider drawing four samples from the 4-dimensional sample space using a PSS 2x2 as shown in Figure \ref{fig:2}. First, we draw four 2-dimensional stratified samples in each subspace $\mathbf{\Theta}_1$ and $\mathbf{\Theta}_2$ (Figure \ref{fig:2}a). The 2-dimensional stratified samples are then randomly paired as shown in Figure \ref{fig:2}b. One could generalize this to cases where the strata are not equally sized in each subdomain but this will not be considered here given the challenges associated with computing sample weights. In particular, this represents the generalization of the Latin hyperrectangle sampling method of Mease and Bingham \cite{Mease_Bingham_Tech_06} in which it is necessary to violate the desirable property that the sum of the probability weights equals one.

The notation used to describe a PSS design identifies, in decreasing dimensional order, the definitions of the subspaces as shown in Figure \ref{fig:4D_PSS}. For low dimensional problems, this can be characterized by simply stating the dimension of all subspaces (e.g.\ PSS$-2\times2$). However, for high dimensional problems, it is convenient to identify the number $K_i$ of $N_i$-dimensional subspaces using a superscript. In general, this is denoted as PSS$-N_1^{K_1}N_2^{K_2}\dots N_{N_S}^{K^{N_S}}$ where $N_1>N_2>\dots>N_{N_S}$. For example, a 100-dimensional problem with 20 4-dimensional subspaces and 5 2-dimensional subspaces is denoted by PSS$-4^{20}2^5$. Note that this alone does not fully specify a PSS. Additionally, it is necessary to stipulate which variables are being grouped in each subspace along with the number of samples and the number of strata in each subspace.

\subsection{Bias and Response Variance}
The response variance from PSS is derived here and shown to follow a nearly identical form to that of LHS. The derivation follows directly from \citep{McKay_et_al_Tech_79} as given below.

Partially Stratified Sampling divides an $N$-dimensional sample space into $N_s$ disjoint subspaces $\mathbf{\Theta}_i$. Each subspace is stratified into regions possessing probability volume $\dfrac{1}{n}$. Considering all possible combinations of strata spanning the subspaces yields $n^{N_s}$ $N$-dimensional cells each having probability volume $\dfrac{1}{n^{N_s}}$. Each cell can be labeled by a set of $N_s$ cell coordinates $\mathbf{m}_i=\{m_{i1},m_{i2},\dots,m_{iN_s}\}$ where $m_{ij}$ is the stratum index of cell $i$ in subspace $\mathbf{\Theta}_j$. A Partially stratified sample of size $n$ is obtained by randomly selecting $n$ of the cells $\mathbf{m}_i; i=1,\dots,n$ such that for each subspace $\mathbf{\Theta}_j$, the set $\{m_{ij}\}_{i=1}^n$ is a random permutation of the integers $1,\dots,n$. A random sample is generated in each selected cell. 

The probability density function of $\mathbf{X}$ given $\mathbf{X}\in$ cell $i$ equals $n^{N_s}f(x)$ if $\mathbf{x}\in$ cell $i$ and zero otherwise. The marginal distribution of $Y_i$ generated from a PSS is given by:
\begin{equation}
\begin{aligned}
P[Y_i\le y] &= \sum_{\substack{\text{cells } \\ q=1}}^{n^{N_s}} P[Y_i\le y|\mathbf{X}\in\text{cell }q]\cdot P[\mathbf{X}\in\text{cell }q] \\
& = \sum_{\substack{\text{cells } \\ q=1}}^{n^{N_s}} \int_{\substack{\text{cell } q \\ h(\mathbf{x})\le y}}n^{N_s}f(\mathbf{x})d\mathbf{x}\cdot n^{-N_s} \\
& = \int_{h(\mathbf{x})\le y}f(\mathbf{x})d\mathbf{x}
\end{aligned}
\end{equation}
which is clearly equal to the distribution of $Y_i$ generated from SRS. Thus, a statistical estimator, $T_P$, constructed from a PSS is an unbiased estimator of $\tau$.

To derive the form of the variance of $T_P$, define the indicator variable $\alpha_i$ as:
\begin{equation}
\alpha_i= 
\begin{cases}
    1,& \text{if } \text{cell }i \text{ is in the sample}\\
    0,              & \text{otherwise}
\end{cases}
\end{equation}
The PSS estimator can be written as:
\begin{equation}
T_P=\dfrac{1}{n}\sum_{i=1}^{n^{N_s}}\alpha_ig(Y_i)
\end{equation}
where $Y_i=h(\mathbf{X}_i)$ and $\mathbf{X}_i\in\text{cell }i$. The variance of $T_P$ is given by:
\begin{equation}
\label{eqn:Var_TP}
\text{Var}[T_P]=\dfrac{1}{n^2}\sum_{i=1}^{n^{N_s}}\text{Var}[\alpha_ig(Y_i)] + \dfrac{1}{n^2}\sum_{i=1}^{n^{N_s}}\sum_{\substack{j=1\\j\ne i}}^{n^{N_s}}\text{Cov}[\alpha_ig(Y_i),\alpha_jg(Y_j)]
\end{equation}
Expanding the first term of Eq.\ \eqref{eqn:Var_TP} yields:
\begin{equation}
\text{Var}[\alpha_ig(Y_i)]=E[\alpha_i^2]\text{Var}[g(Y_i)]+E[g(Y_i)]^2\text{Var}[\alpha_i]
\end{equation}
Exploiting the fact that $E[\alpha_i]=E[\alpha_i^2]=\dfrac{1}{n^{{N_s}-1}}$ yields:
\begin{equation}
\label{eqn:sum_var1}
\sum_{i=1}^{n^{N_s}}\text{Var}[\alpha_ig(Y_i)]=\dfrac{1}{n^{{N_s}-1}}\sum_{i=1}^{n^{N_s}}E[g(Y_i)-\mu_i]^2+\left(\dfrac{1}{n^{{N_s}-1}}-\dfrac{1}{n^{2{N_s}-2}}\right)\sum_{i=1}^{n^{N_s}}\mu_i^2
\end{equation}
where $\mu_i=E[g(Y)|\mathbf{X}\in\text{cell }i]$. By definition:
\begin{equation}
E[g(y_i)-\mu_i]^2=n^{N_s}\int_{\text{cell }i}(g(y)-\tau)^2f(x)dx+(\mu_i-\tau)^2
\end{equation}
which reduces Eq.\ \eqref{eqn:sum_var1} to:
\begin{equation}
\label{eqn:sum_var2}
\sum_{i=1}^{n^{N_s}}\text{Var}[\alpha_ig(Y_i)]=n\text{Var}[Y]-\dfrac{1}{n^{{N_s}-1}}\sum_{i=1}^{n^{N_s}}(\mu_i-\tau)^2+\left(\dfrac{1}{n^{{N_s}-1}}-\dfrac{1}{n^{2{N_s}-2}}\right)\sum_{i=1}^{n^{N_s}}\mu_i^2
\end{equation}
Consider now the covariance term in Eq.\ \eqref{eqn:Var_TP}, which can be expanded as:
\begin{equation}
\label{eqn:sum_cov}
\sum_{i=1}^{n^{N_s}}\sum_{\substack{j=1\\j\ne i}}^{n^{N_s}}\text{Cov}[\alpha_ig(Y_i),\alpha_jg(Y_j)]=\sum_{i=1}^{n^{N_s}}\sum_{\substack{j=1\\j\ne i}}^{n^{N_s}}\mu_i\mu_jE[\alpha_i\alpha_j]-\dfrac{1}{n^{2{N_s}-2}}\sum_{\substack{j=1\\j\ne i}}^{n^{N_s}}\mu_i\mu_j
\end{equation}
The value of $E[\alpha_i\alpha_j]$ depends on the cell coordinates $\mathbf{m}_i$. If cell $i$ and cell $j$ have no cell coordinates in common then:
\begin{equation}
\label{eqn:Ewiwj1}
\begin{aligned}
E[\alpha_i\alpha_j]& = E[\alpha_i\alpha_j|\alpha_j=0]P[\alpha_j=0]+E[\alpha_i\alpha_j|\alpha_j=1]P[\alpha_j=1]\\
& =\dfrac{1}{(n(n-1))^{{N_s}-1}}
\end{aligned}
\end{equation}
The case where cell $i$ and cell $j$ share at least one common cell coordinate corresponds to an inadmissible case (i.e.\ $\alpha_i$ and $\alpha_j$ cannot both equal 1) and thus: 
\begin{equation}
\label{eqn:Ewiwj2}
E[\alpha_i\alpha_j]=0.
\end{equation}
Combining Eqs.\ \eqref{eqn:sum_var2} and \eqref{eqn:sum_cov} and utilizing the relations in Eqs.\ \eqref{eqn:Ewiwj1} and \eqref{eqn:Ewiwj2} gives the following relation for the variance of a statistical estimator using PSS:
\begin{multline}
\label{eqn:Var_TP2}
\text{Var}[T_P]=\dfrac{1}{n}\text{Var}[Y]-\dfrac{1}{n^{{N_s}+1}}\sum_{i=1}^{n^{N_s}}(\mu_i-\tau)^2+\left(\dfrac{1}{n^{{N_s}+1}}-\dfrac{1}{n^{2{N_s}}}\right)\sum_{i=1}^{n^{N_s}}\mu_i^2+\\
+\dfrac{n^{{N_s}-1}}{(n-1)^{{N_s}-1}}\sum\sum_R\mu_i\mu_j-\dfrac{1}{n^{2{N_s}}}\sum_{i=1}^{n^{N_s}}\sum_{\substack{j=1\\j\ne i}}^{n^{N_s}}\mu_i\mu_j
\end{multline}
where $R$ denotes the space of $n^{N_s}(n-1)^{N_s}$ admissible cell pairs with cell mean values $(\mu_i,\mu_j)$. Finally, utilizing $\sum\mu_i=n^{N_s}\tau$ and simple algebra, Eq.\ \eqref{eqn:Var_TP2} reduces to:
\begin{equation}
\text{Var}[T_P]=\text{Var}[T_R]+\dfrac{n-1}{n}\dfrac{1}{n^{N_s}(n-1)^{N_s}}\sum\sum_R(\mu_i-\tau)(\mu_j-\tau).
\end{equation}

As previously stated, the variance derived above follows directly from \citep{McKay_et_al_Tech_79} with only slight modifications based on the use of ${N_s}$ stratified subspaces rather than $N$ stratified lines. It follows precisely the same form and is straightforward to show that, when ${N_s}=N$, each stratified subspace corresponds to a component of $\mathbf{X}$ resulting in a Latin hypercube sample. Thus, as in LHS, the proposed PSS method produces a variance reduction over SRS only when:
\begin{equation}
\dfrac{n-1}{n}\dfrac{1}{n^{N_s}(n-1)^{N_s}}\sum\sum_R(\mu_i-\tau)(\mu_j-\tau)\le0.
\end{equation}
In other words, a variance reduction is achieved when the covariance between cells having no common cell coordinates is negative.

\subsection{Asymptotic Properties}
\label{sec:properties}
Stein \cite{Stein_Tech_87} provides proofs of several important asymptotic properties of Latin hypercube samples. Most notably, he shows that, in the limit $n\to\infty$, statistical estimates from Latin hypercube samples are normal possessing variance that is at least as small as an estimate using SRS. Furthermore, he proves that Latin hypercube sampling has the effect of filtering the additive components of $h(\mathbf{X})$. In this section, we generalize these asymptotic properties to partially stratified samples.

Consider the general transformation $h(\mathbf{X})$. By writing the transformation in the form:
\begin{equation}
\label{eqn:best_additive}
h(\mathbf{X}) = h_a(\mathbf{X})+r(\mathbf{X})
\end{equation}
where $h_a(\mathbf{X})$ is the best additive fit to $h(\mathbf{X})$ having form:
\begin{equation}
h_a(\mathbf{X})=\sum_{i=1}^Ng_i(x_i)
\end{equation}
and $r(\mathbf{X})$ is a non-additive remainder function describing variable interactions, Stein \cite{Stein_Tech_87} shows that Latin hypercube sampling filters out the contribution $h_a(\mathbf{X})$. Thus, the primary contributor to the variance of the estimate is the interaction function $r(\mathbf{X})$. 

We submit the following alternative form for $h(\mathbf{X})$:
\begin{equation}
\label{eqn:alt_form}
h(\mathbf{X})=f(\mathbf{X})+r(\mathbf{X})
\end{equation}
where $f(\mathbf{X})$ takes instead the form:
\begin{equation}
f(\mathbf{X})=\sum_{i=1}^{N_s}f_i(\mathbf{X}_i)
\label{eqn:functional}
\end{equation}
such that $\cup_i\mathbf{X}_i=\mathbf{X}$ and $\cap_i\mathbf{X}_i=\emptyset$. In other words, instead of decomposing $h(\mathbf{X})$ into an additive sum of functions on a single variable, we decompose it into an additive sum of multivariate functionals. These multivariate functionals may or may not be additive - or even close to additive. If functional $f_i(\mathbf{X}_i)$ is nearly additive, then it may be approximated as:
\begin{equation}
\label{eqn:sub_additive}
f_i(\mathbf{X}_i)\approx \sum_{k=1}^{N_i}g_k(x_k)
\end{equation}
and Latin hypercube sampling of this subspace is appropriate to filter out the contributions of $g_k(x_k)$. When all such $f_i(\mathbf{X}_i) \hspace{3pt}\forall i$ are closely approximated by the form in Eq.\ \eqref{eqn:sub_additive} then Latin hypercube sampling of the full space will be effective. However, if \emph{any} $f_i(\mathbf{X}_i)$ is not well-approximated by Eq.\ \eqref{eqn:sub_additive}, alternative sampling methods should be considered on that subspace. In particular, if there are strong interaction terms it will be demonstrated that a stratified design of the subspace is more effective than a Latin hypercube for low-dimensional subspaces.

An alternative presentation that leads to the same conclusion follows. Consider an $N$-dimensional PSS possessing $N_s$ orthogonal subspaces $\mathbf{\Theta}_i; i=1,\dots,N_s$ each having dimension $N_i$ such that $\sum_{i=1}^{N_s}N_i=N$. For each subspace, define a new variable as follows:
\begin{equation}
Z_i=f_i(\mathbf{X}_i)
\end{equation}
where $\mathbf{X}_i$ is the vector of variables spanning subspace $\mathbf{\Theta}_i$, such that we can write $Y=h(\mathbf{X})\approx F(\mathbf{Z})$. In general, the functional forms of $f_i(\cdot)$ and $F(\cdot)$ may be difficult or even impossible to determine. But, defining these variables conveniently reduces each subspace to a single dimension; effectively transforming a PSS on $\mathbf{X}$ into a Latin hypercube sample on the lower $N_s$-dimensional vector $\mathbf{Z}$. 

Next, consider two cases: additive $Z_i$ and non-additive $Z_i$. If $Z_i$ is additive then:
\begin{equation}
\label{eqn:additive_sub}
Z_i=f_i(\mathbf{X}_i)=\sum_{k=1}^{N_i}f_i(X_k)
\end{equation}
In this case, the most sensible approach is to perform Latin hypercube sampling on $\mathbf{X}_i$ - which is the same as defining $N_i$ 1-dimensional subspaces as opposed to a single $N_i$-dimensional subspace. In other words, higher-dimensional subspaces should not be applied to these variables.

If, on the other hand, $Z_i$ is non-additive then performing LHS on $\mathbf{X}_i$ is not necessarily the best approach. Consider the following non-additive form for the variables $Z_i$:
\begin{equation}
\label{eqn:non-additive}
Z_i=\prod_{k=1}^{N_i}\gamma_kX_k^{\beta_k}
\end{equation}
where the subscript $k$ refers to each dimension of the subspace and $\gamma_k$ and $\beta_k$ are defined arbitrarily. Clearly, $Z_i$ are non-additive in $X_k$ but, it should be emphasized that this is not a general form for non-additive functions. It is used purely for convenience of demonstration. As will be seen, a true stratified design provides superior variance reduction on variable $Z_i$ when the dimension of $\mathbf{X}_i$ is low. 

Considering that a PSS can be essentially reduced to a Latin hypercube on the variables $Z_i$, it follows that the asymptotic properties of partially stratified sampling can be inferred directly from those of Latin hypercube sampling given by Stein \cite{Stein_Tech_87}. First, asymptotic normality is preserved ($Z-i$ are independent) - as is the property that a PSS will not increase the variance of a statistical estimator over SRS given a large enough $N$. Moreover, PSS has the effect of filtering out additive combinations of non-additive functions. In other words, a partially stratified design filters out the additive components in $Z_i$. This is an important result because it states that, for a general non-additive function that can be written as an additive combination of simpler non-additive functions, the variance can be further reduced from a LHS by improving the sampling method specifically on the non-additive components. This will be demonstrated to have significant effect on variance reduction in the following sections.


\section{When to use partially stratified sampling?}
In this section, we explore the use of PSS. First, we discuss the benefits of true stratified sampling for low dimensional problems with strong interactions - showing that it produces a variance reduction that is superior to LHS and indicating that SS on low-dimensional subspaces where variables interact is likely to be preferable. Next, we explore the relationship between interaction strength and its influence on the decision to perform true stratified sampling or LHS on a subdomain. Lastly, we identify and discuss the primary challenges associated with selecting and implementing a PSS design.

\subsection{Benefits of true stratified sampling in low dimension} 

True stratified sampling has the benefit of improved variance reduction for low dimensional transformations possessing  interaction terms as recently demonstrated by Shields et al. \cite{Shields_et_al_RESS_15}. As illustrated in Eqs.\ \eqref{eqn:ss_var} and \eqref{eqn:lhs_var}, SS and LHS reduce variance through different statistical mechanisms. SS reduces variance by considering the global statistical estimate to be the statistical mean of a set of local means conditioned upon the strata. In this way, variance is reduced for the main and interactive effects in equal measure. LHS, on the other hand, reduces variance by creating negative covariance between the transformed samples. As Stein shows, the strength of the covariance directly relates to the strength of the main effects (see  \cite{Stein_Tech_87}, Theorem 1). Thus, the stronger the main effects, the greater the variance reduction. However, for transformations possessing interactions, the variance reduction may be less than that achieved by SS. This is demonstrated clearly by the following simple example. 

\begin{figure}
\centering
\subfigure[\label{fig:additive}]{
\centering
\includegraphics[width=0.47\columnwidth]{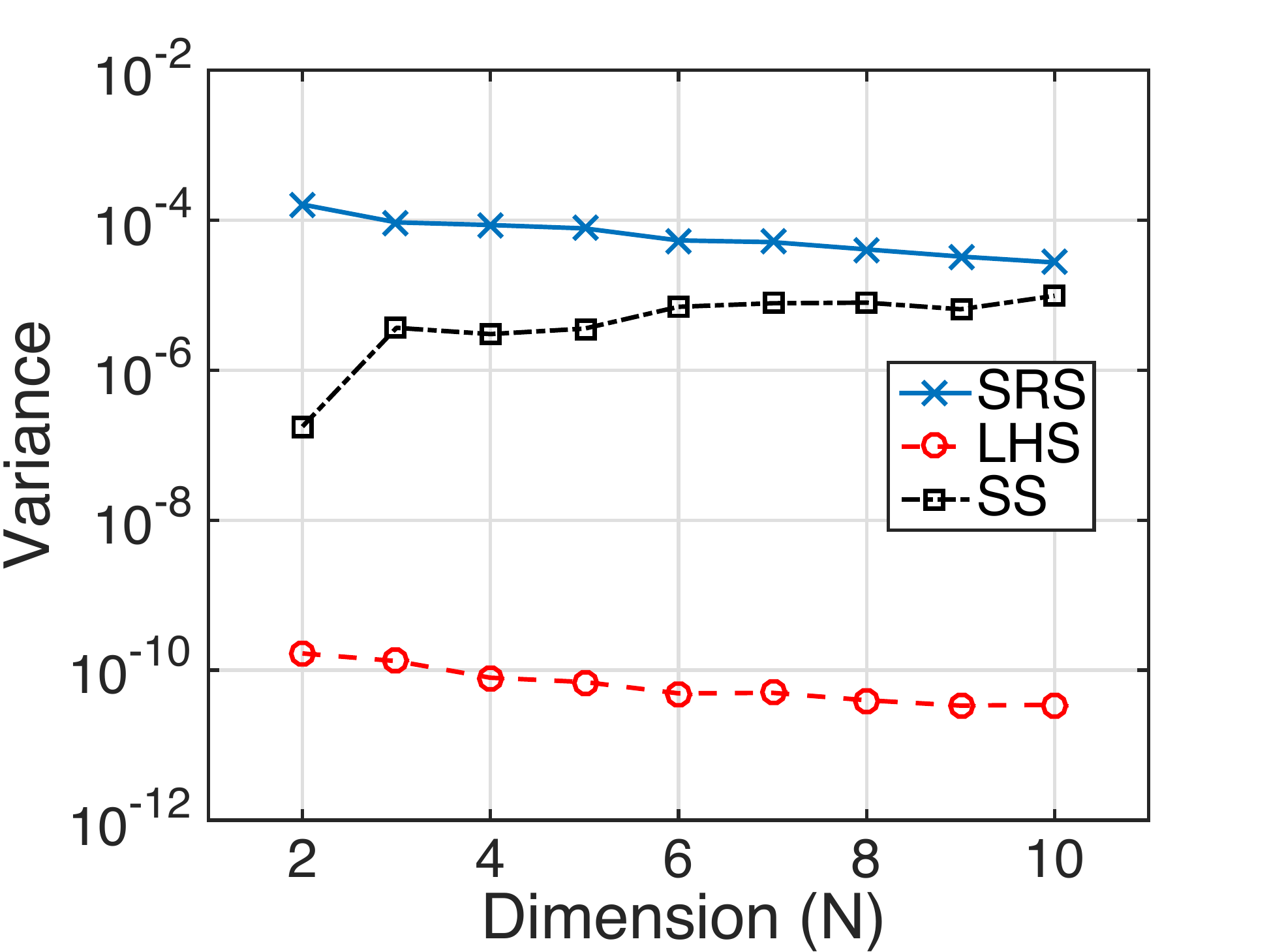}
}
\subfigure[\label{fig:product}]{
\centering
\includegraphics[width=0.47\columnwidth]{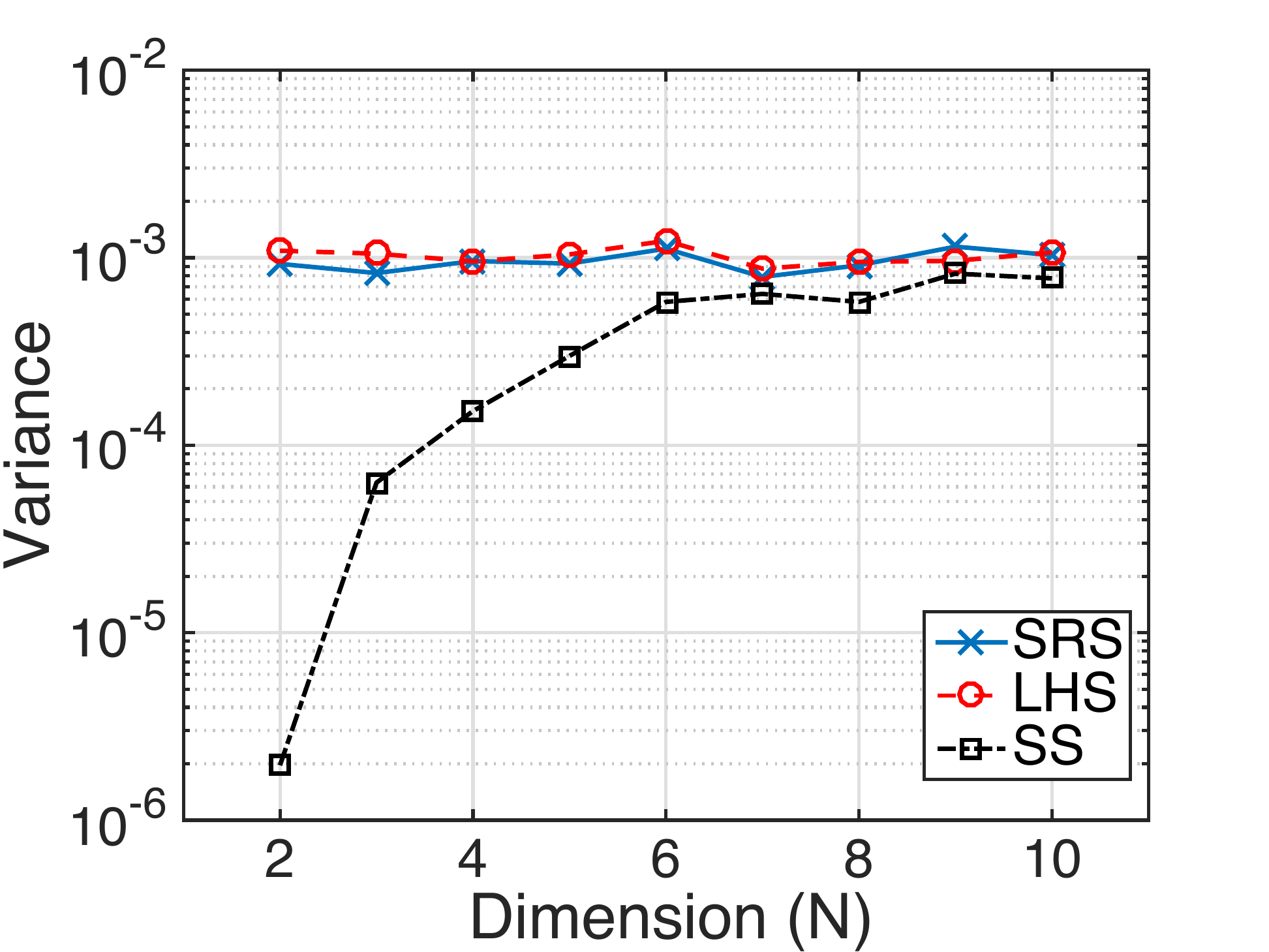}
}
\caption{Variance of Monte Carlo estimates of the mean value for (a) an additive function and (b) a product function using stratified sampling (SS), Latin hypercube sampling (LHS) and simple random sampling (SRS). }
\label{fig:3}
\end{figure}

Consider two simple transformations. The first is an additive function defined by:
\begin{equation}
Y_1 = \dfrac{2}{N}\sum_{i=1}^NX_i
\end{equation}
where $X_i\sim U(0,1)$ such that $E[Y_1]=1$. The second is a product function defined by:
\begin{equation}
Y_2 = \prod_{i=1}^NX_i
\label{eqn:prod}
\end{equation}
where $X_i\sim U\left(-\sqrt{3},\sqrt{3}\right)$ such that $E[Y_2]=0$. Note that, when $E[X_i]\ne0$, $Y_2$ will also possess noticeable main effects although, as shown in \cite{Shields_et_al_RESS_15}, stratified sampling is still shown to reduce variance more than LHS in this case. In both functions, $X_i$ are iid so it follows that $Y_1$ and $Y_2$ are asymptotically normal and lognormal respectively as $N\to\infty$. Figure \ref{fig:3} shows the variance of estimates for $E[Y_1]$ and $E[Y_2]$ from 1,024 samples generated using SRS, LHS, and SS for different values of $N$. 
For $Y_1$, LHS produces by far the best estimates of $\mu_{Y_1}$ (as expected based on Stein's findings) yielding variance that is effectively negligible ($\sim 10^{-10}$). This is reduced by more than four orders of magnitude over SS, which yields still significant variance reduction over SRS. The variance reduction improves slowly for LHS with the dimension $N$ but the variance of SS increases slowly with dimension given the inability to simultaneously stratify many dimensions. 

The product function, $Y_2$, however tells a different story. For low-dimensional problems, SS produces a significant reduction in variance compared to LHS (which, as expected from Stein \cite{Stein_Tech_87} yields no variance reduction at all) that is up to three orders of magnitude. The reduction in variance over SRS/LHS diminishes with dimension although it maintains some minor advantage through moderate dimensions ($N=10$). The performance loss results from the increase in variance of the strata means $\mu_i$ as the dimension grows. More specifically, although stratum probability $p_i$ remains constant, the extents of the strata in each direction grow as the dimension grows. For example, given 1,024 samples and strata probabilities $p_i=\dfrac{1}{1024}\forall i$, the stratum extents in a 2-dimensional space are $\dfrac{1}{32}$ ($\left(\dfrac{1}{32}\right)^2=\dfrac{1}{1024}=p_i$) while the corresponding extents in a 10-dimensional space are $\dfrac{1}{2}$ ($\left(\dfrac{1}{2}\right)^{10}=\dfrac{1}{1024}=p_i$). 

These results make a compelling case for the use of PSS. For a general, high dimensional transformation possessing both main effects and interactions it will be advantageous to isolate the main effects and interaction effects to the largest extent possible and sample them in different ways (LHS for main effects and SS for interactions). The PSS methodology provides this capability.

\subsection{Effect of interaction strength}

The analysis above presents a difficult question: When is it appropriate to use true SS on a subspace and when should LHS be used? To investigate this question, we employ global or variance-based sensitivity analysis \cite{Saltelli_et_al_08} to make a connection between interaction strength and the the choice of sampling method for a given subdomain. For this purpose, let us define the main effects of the transformation through the conditional variances:
\begin{equation}
V_i=\text{Var}\{E[Y|X_i]\}
\end{equation}
and the interaction effects through the conditional variances:
\begin{equation}
V_{ij}=\text{Var}\{E[Y|X_i,X_j]\}.
\end{equation}
Using this notation, we can write the total variance $\text{Var}\{Y\}$ using the ANOVA-HDMR (Analysis of Variance - High Dimensional Model Representation) if $\mathbf{X}$ are independent variables as \cite{Saltelli_et_al_08}:
\begin{equation}
V_T=\sum_iV_i+\sum_i\sum_{j>i}V_{ij}+\dots+V_{12\dots k}
\end{equation}
Such decomposition of variance is common in global sensitivity analysis where the sensitivity indices (Sobol indices) $S_i=\dfrac{V_i}{V_T}$, $S_{ij}=\dfrac{V_{ij}}{V_T}$, etc.\ provide normalized contribution factors for the main effects and interaction effects respectively \cite{Saltelli_et_al_08}. 

Consider now the functional $f_i(\mathbf{X}_i)$ from Eq.\ \eqref{eqn:functional} and its decomposition into main effects and interactions as:
\begin{equation}
f_i(\mathbf{X}_i) = f_{ia}(\mathbf{X})+r_i(\mathbf{X})
\label{eqn:add_functional}
\end{equation}
where $f_{ia}(\mathbf{X})$ and $r_i(\mathbf{X})$ are the main and interactive contributions respectively. As a general rule, Latin hypercube sampling will produce significant variance reduction on $f_{ia}(\mathbf{X})$ and no variance reduction on $r_i(\mathbf{X})$. As shown by Stein \cite{Stein_Tech_87}, the variance of the LHS estimator $T_{Li}$ on this subdomain can be expressed as:
\begin{equation}
\text{Var}[T_{Li}]=\dfrac{1}{n}\int r_i(\mathbf{x})^2dF_i(\mathbf{x})+O(n^{-1})
\label{eqn:lhs_var2}
\end{equation}
which equates to standard Monte Carlo sampling on the interaction functional $r_i(\mathbf{X})$. Utilizing the ANOVA-HDMR decomposition, it follows that:
\begin{equation}
\int r_i(\mathbf{x})^2dF_i(\mathbf{x})=\sum_i\sum_{j>i}V_{ij}+\dots+V_{12\dots k}
\label{eqn:lhs_var_dec}
\end{equation}
yielding:
\begin{equation}
\text{Var}[T_{Li}]=\dfrac{1}{n}\left[\sum_i\sum_{j>i}V_{ij}+\dots+V_{12\dots k}\right]+O(n^{-1})
\end{equation}
such that the performance of LHS on a given subdomain can be directly related to the interaction sensitivities. 

\begin{figure}[!ht]
\centering
\includegraphics[width=0.5\columnwidth]{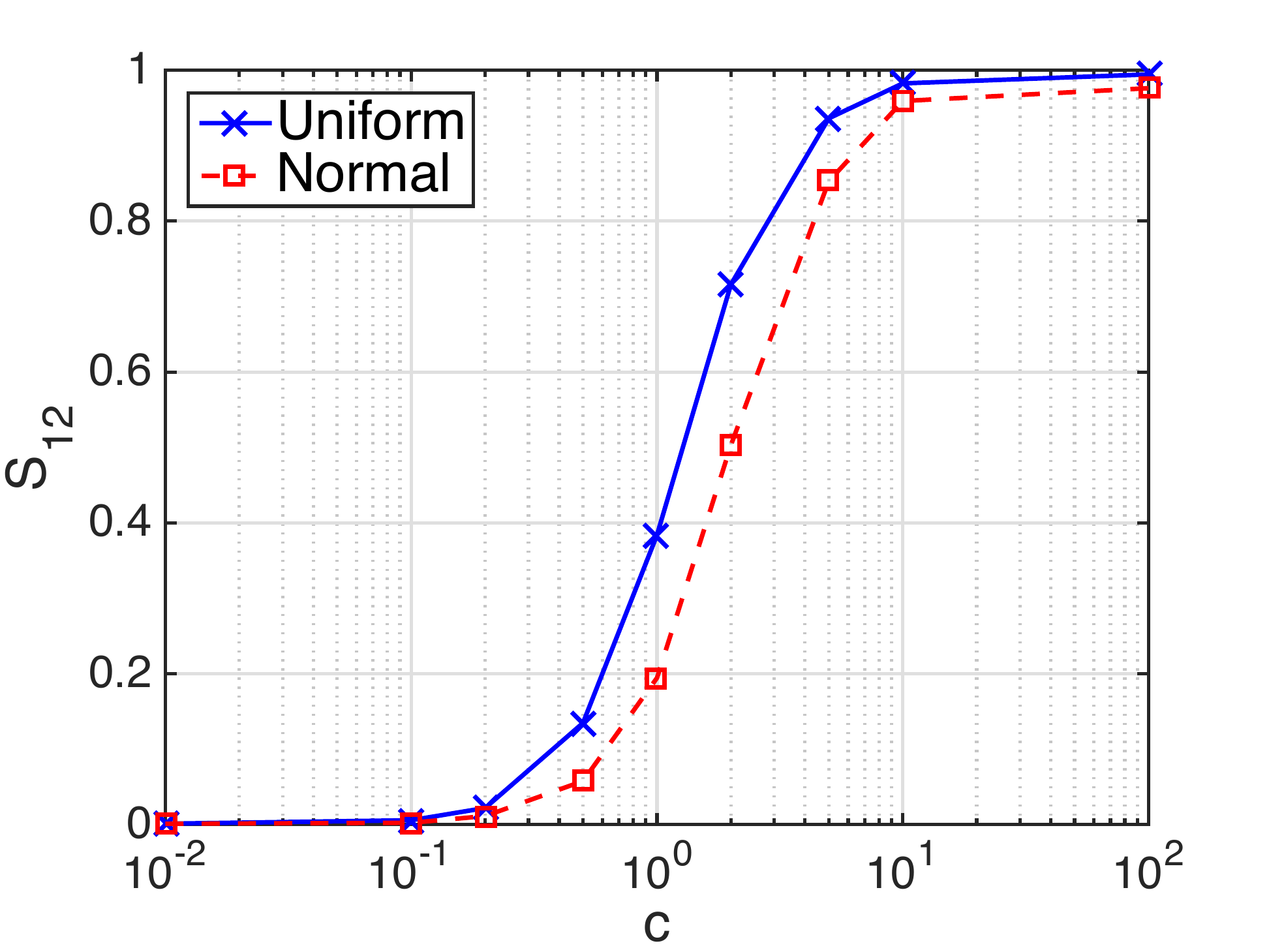}
\caption{Dependence of interaction sensitivity on the interaction coefficient, $c$, for a second order polynomial. See Eq.\ \eqref{eqn:2D_square}.}
\label{fig:4}
\end{figure}

Having established that true SS is more effective than LHS at reducing variance associated with interactions but is much less effective at reducing variance associated with main effects, we can make the qualitative statement that true SS is appropriate on a given subdomain when the contribution of the main effects to the variance from SS is smaller than the variance from LHS resulting from interactions. To elucidate this point, consider the simple example of a 2D quadratic functional with interactions:
\begin{equation}
\label{eqn:2D_square}
Z_i=f_i(\mathbf{X}_i) = X_1^2+X_2^2+cX_1X_2
\end{equation}
such that it is reasonable to assume $f_{ia}(\mathbf{X})=X_1^2+X_2^2$ and $r_i(\mathbf{X})=cX_1X_2$ when $E[X_1]=E[X_2]=0$ (note the product $X_1X_2$ may have considerable main effects when $E[X_1],E[X_2] \ne 0$). We consider two different distributions for $\mathbf{X}$ such that $\mathbf{X}\sim N(0,1)$ and $\mathbf{X}\sim U(-\sqrt{3},\sqrt{3})$ and study the effect of the coefficient $c$ on the variance of the estimate $\mu_{Z_i}$ of the mean $E[Z_i]$. For the two distributions, Figure \ref{fig:4} shows the interaction sensitivity indices computed using a Monte Carlo approach \cite{Saltelli_et_al_08} as a function of $c$ follows a logistic form when $c$ is plotted in log scale. 
This suggests that interaction sensitivity is essentially invariant for small $c$ ($c<0.1$ corresponds to essentially negligible interaction) and large $c$ ($c>10$ corresponds to essentially negligible main effects) and varies strongly in the range $0.1<c<10$. Figure \ref{fig:4} further shows that $f_i(\mathbf{X}_i)$ is more sensitive to interactions when $\mathbf{X}$ is uniformly distributed. 
\begin{figure}[!ht]
\centering
\subfigure[\label{fig:5a}Normal Distribution]{
\centering
\includegraphics[width=0.46\columnwidth]{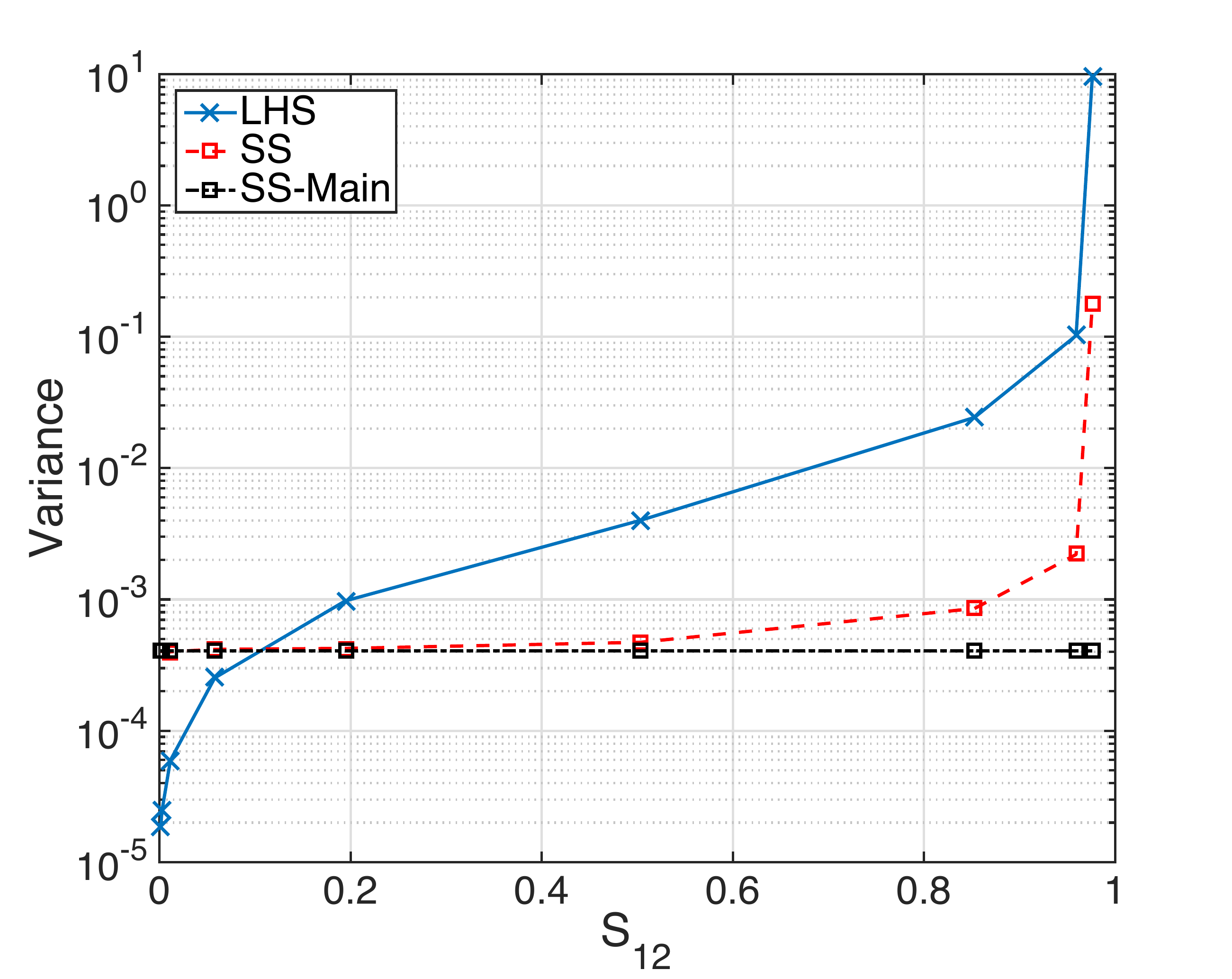}
}
\subfigure[\label{fig:5b}Uniform Distribution]{
\centering
\includegraphics[width=0.46\columnwidth]{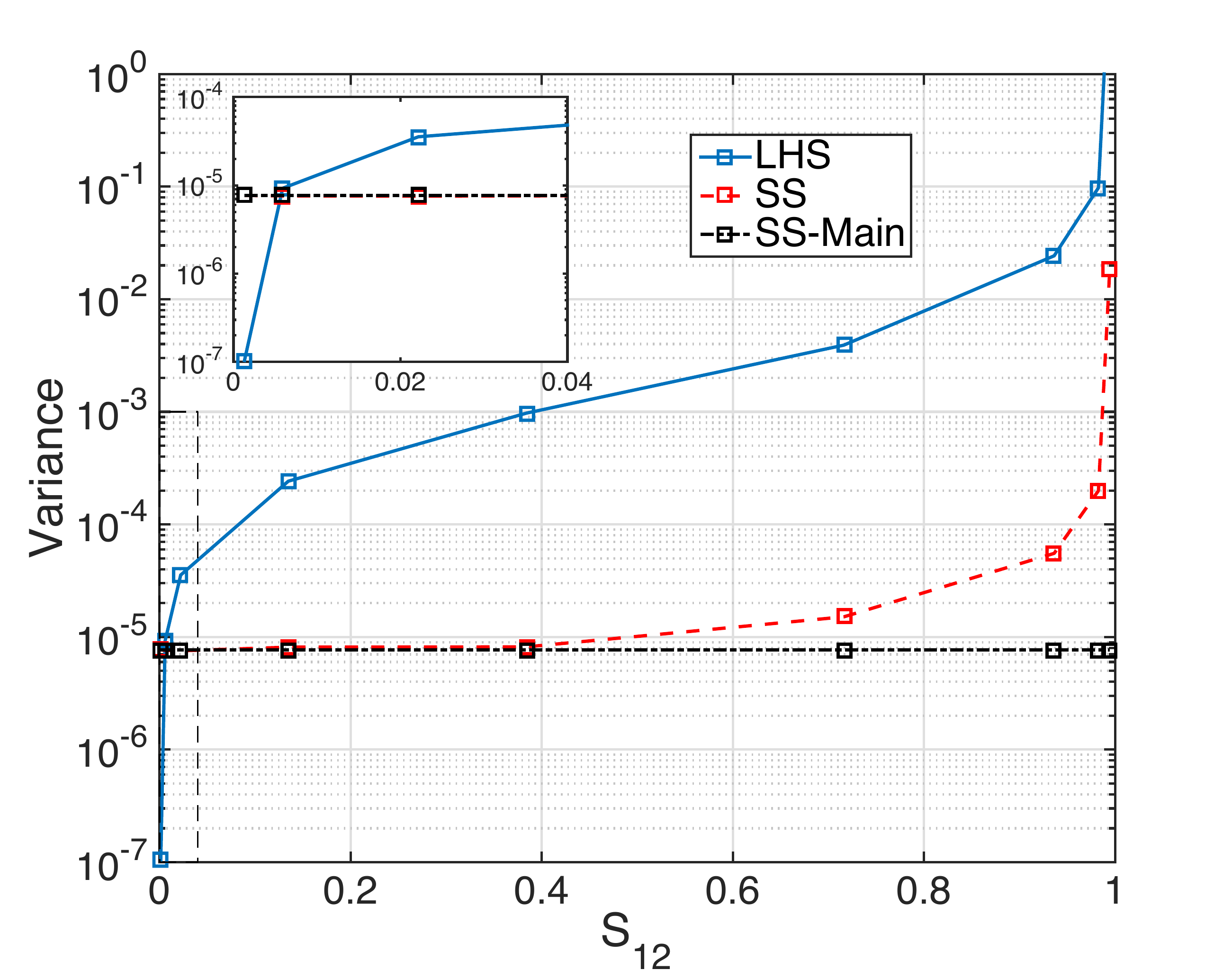}
}
\caption{Dependence of variance on the interaction sensitivity for stratified sampling and Latin hypercube sampling given (a.) normal input variables and (b.) uniform input variables. A very small degree of interaction dramatically increases the variance of LHS estimators relative to true SS.}
\label{fig:5}
\end{figure}
Figure \ref{fig:5} shows the variance of the estimate $\mu_{Z_i}$ from LHS (variance essentially results entirely from interaction effects - see Eqs.\ \eqref{eqn:lhs_var2} and \eqref{eqn:lhs_var_dec}) and true SS as a function of interaction sensitivity $S_{12}$ for the normal distribution (Figure \ref{fig:5a}) and uniform distribution (Figure \ref{fig:5b}). Also shown in Figure \ref{fig:5} is the variance of the stratified estimator attributed to main effects. These plots ratify our previous assertion that true SS is appropriate when the variance of the SS estimator associated with main effects is less than the variance of the LHS estimator associated with interactions effects but also highlights several other interesting and important factors in sample design selection. 1. For true SS, the variance associated with interaction effects is negligible unless the the interactions are very strong ($S_{12}>0.5$); 2. For LHS, the variance of the mean increases dramatically (orders of magnitude) with the introduction of even very small interaction effects. 3. SS becomes superior to LHS at very low sensitivity values ($S_{12}=0.005$ for uniform $\mathbf{X}$ and $S_{12}=0.1$ for normal $\mathbf{X}$). This leads to the conclusion that SS on a subdomain will likely be superior to LHS unless the interaction effects are weak.

%

\subsection{Challenges}
The most obvious and significant challenge to partially stratified sampling is to identify the optimal subspace decomposition. In general, the transformation $h(\mathbf{X})$ is rarely given in closed form and usually takes the form of a numerical computer model possessing complex interactions of variables. In certain cases, it may be clear which variables are interacting which will inform the PSS subspace definitions - but such cases are the exception. This challenge will be addressed in the following section.

A second challenge is that, in order to be effective, the subspaces $\mathbf{\Theta}_i$ must remain relatively low-dimensional. The true SS that occurs on these subspaces may perform poorly in high-dimensional spaces (such that e.g.\ $2^{N_i}>N$). Consequently, it may be difficult to reduce the variances resulting from interactions of many variables as may occur if, for example, a component $Z_i$ in Eq.\ \eqref{eqn:non-additive} is defined by the product of a large number of input variables $X_i$.


\section{``Latinized" Stratified Sampling}
\begin{figure}[!ht]
\centering
\includegraphics[width=0.3\columnwidth]{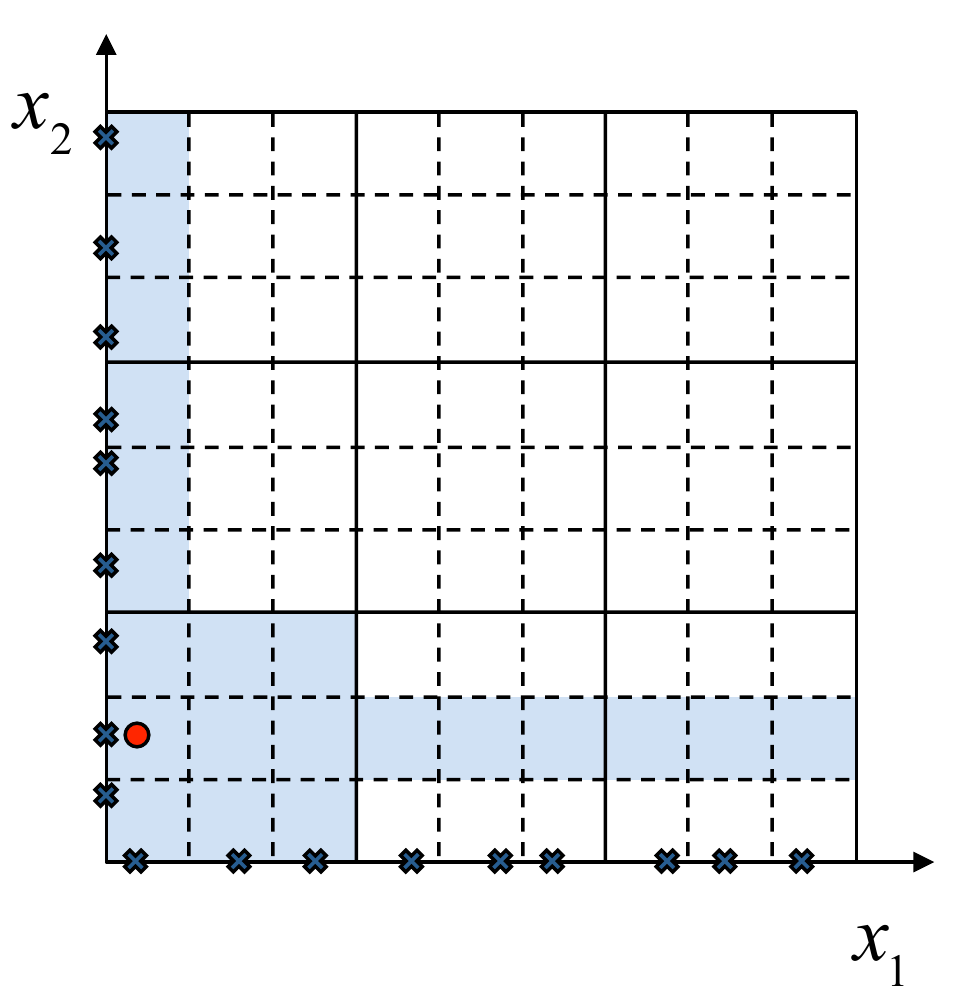}
\includegraphics[width=0.3\columnwidth]{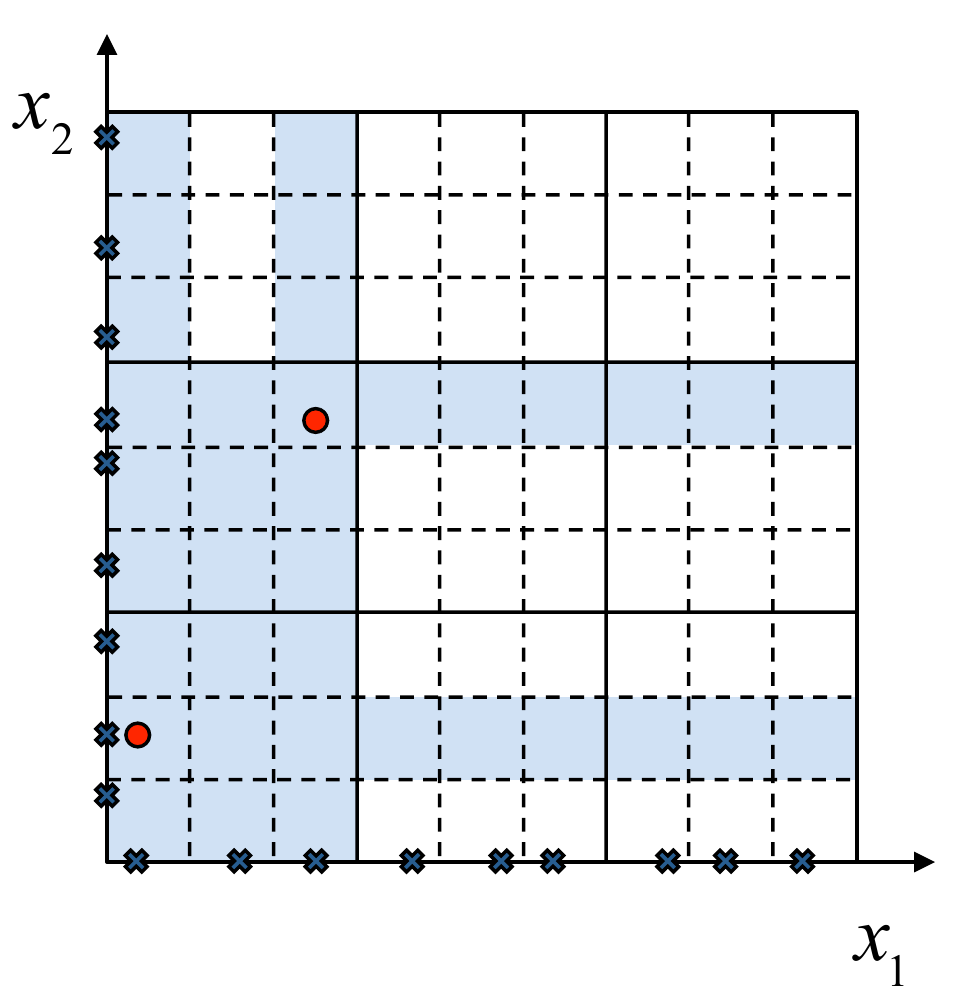}
\includegraphics[width=0.3\columnwidth]{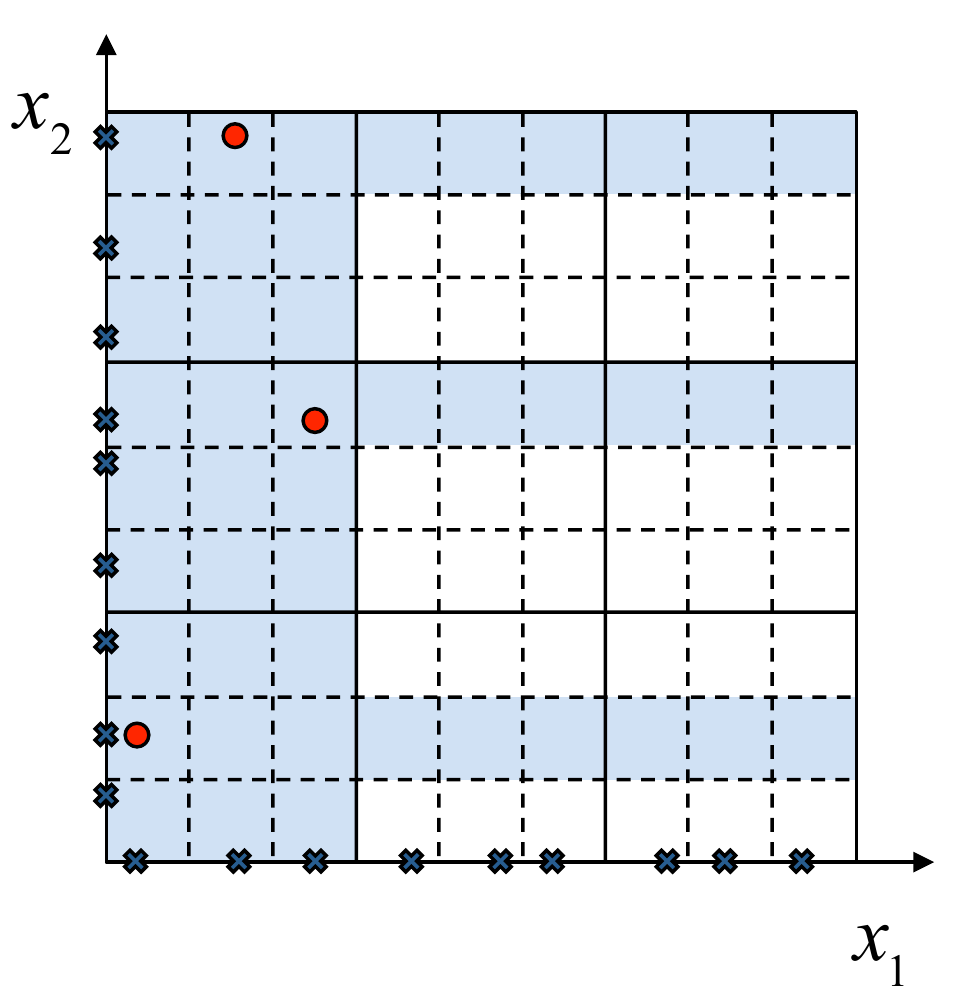}
\includegraphics[width=0.3\columnwidth]{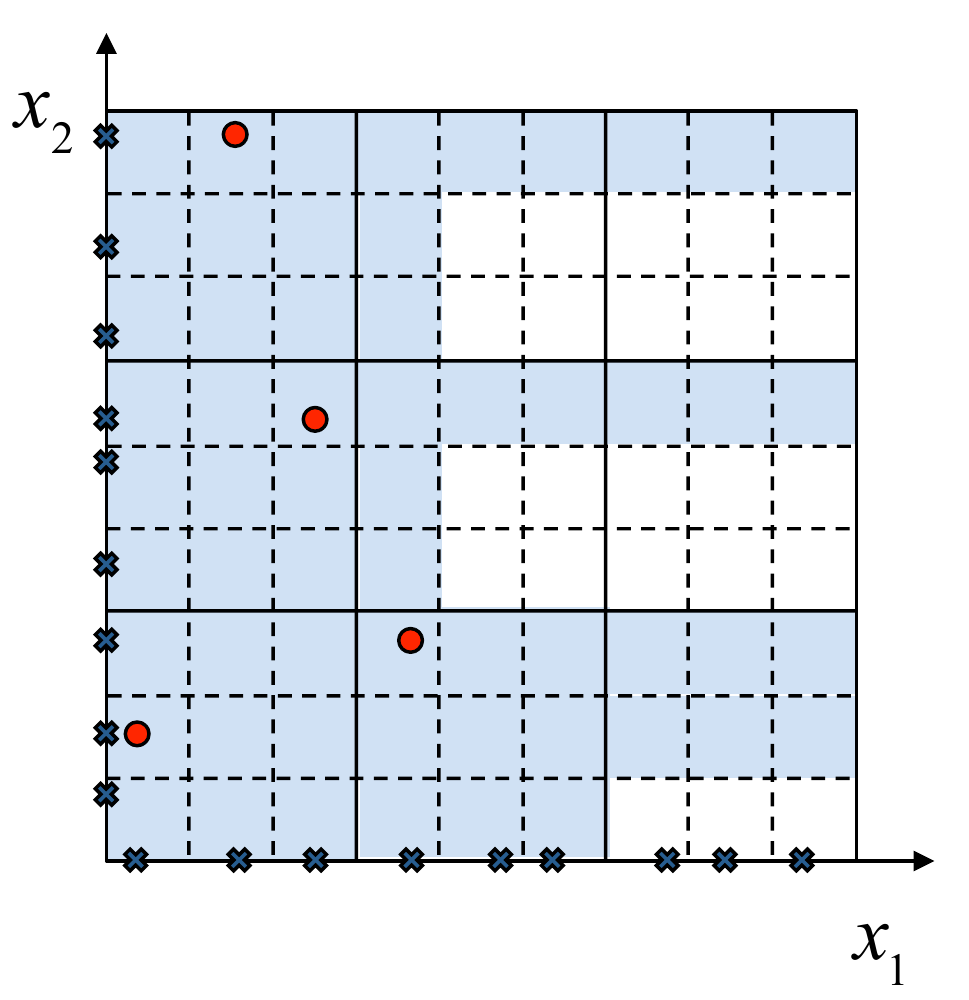}
\includegraphics[width=0.3\columnwidth]{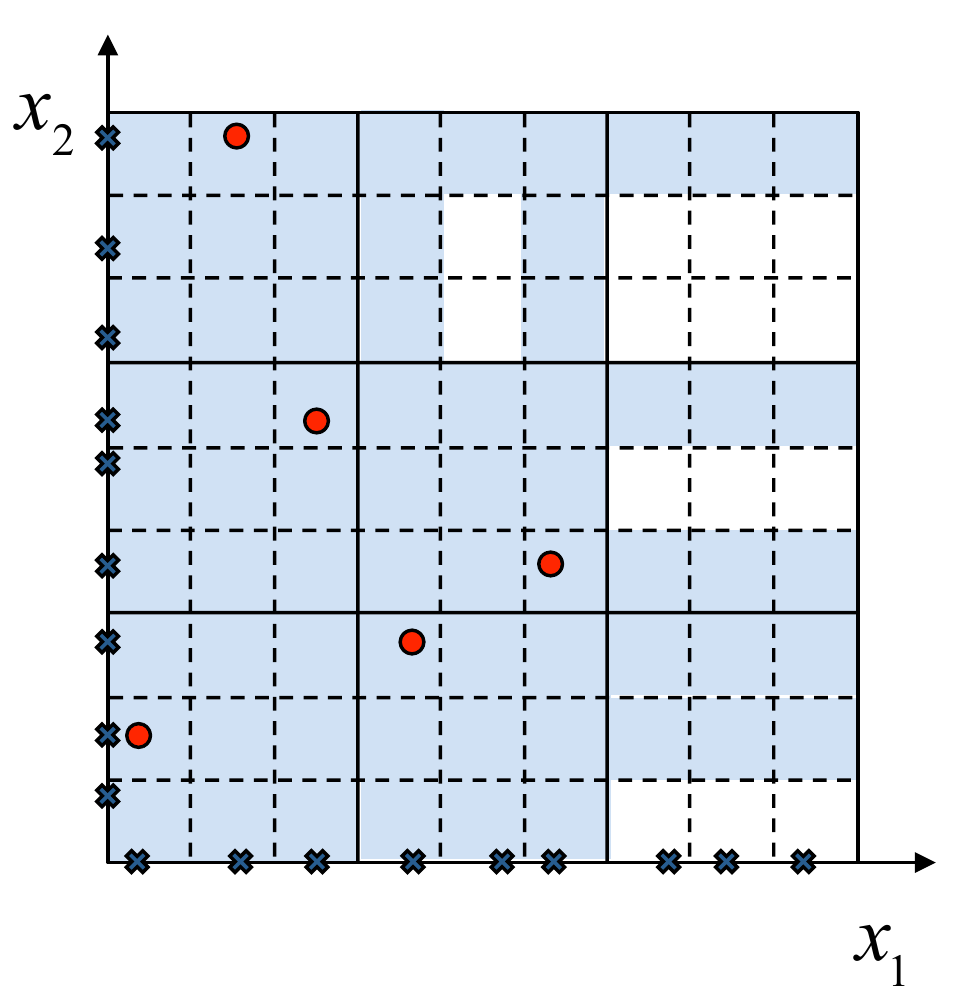}
\includegraphics[width=0.3\columnwidth]{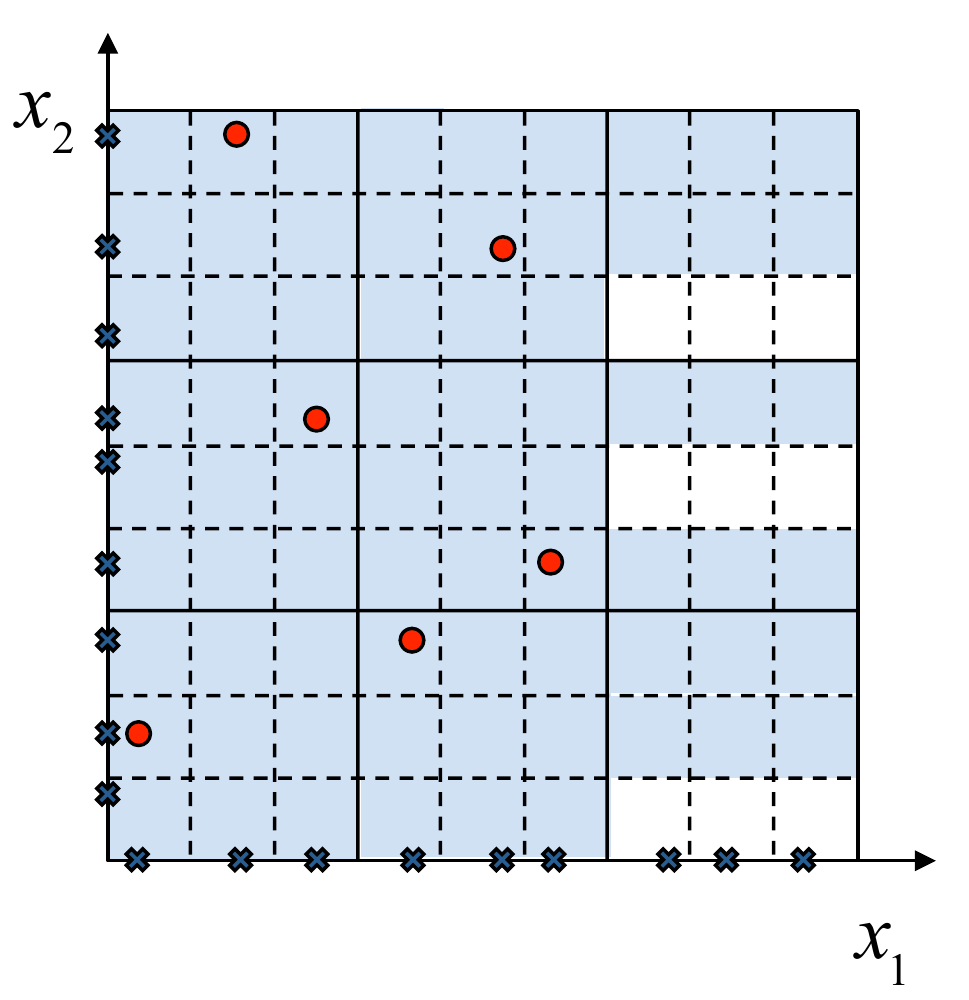}
\includegraphics[width=0.3\columnwidth]{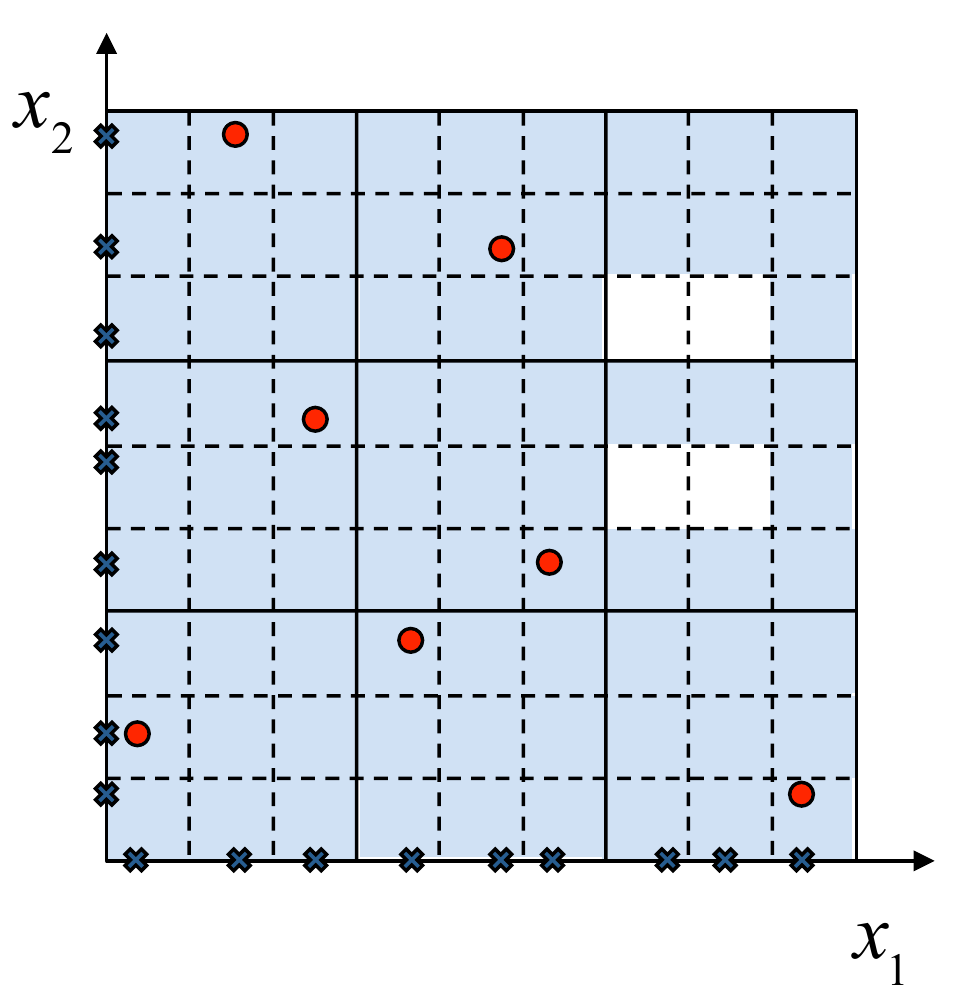}
\includegraphics[width=0.3\columnwidth]{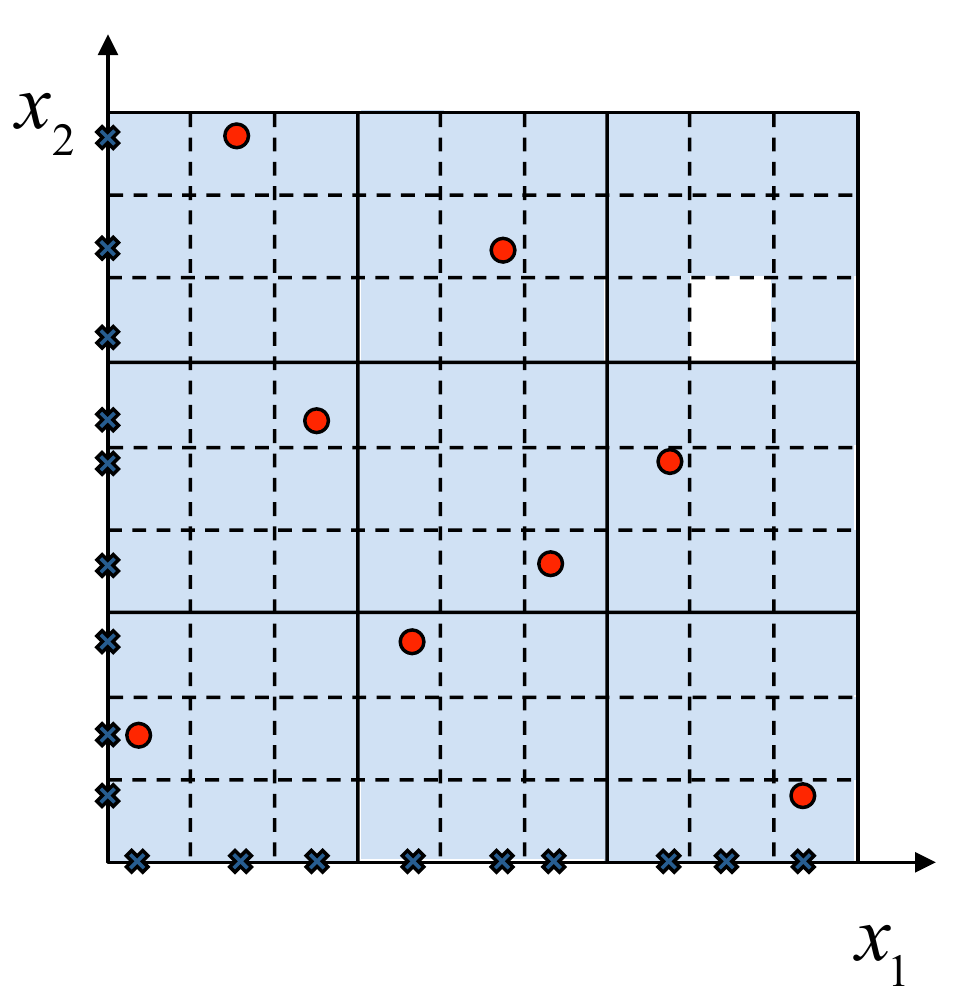}
\includegraphics[width=0.3\columnwidth]{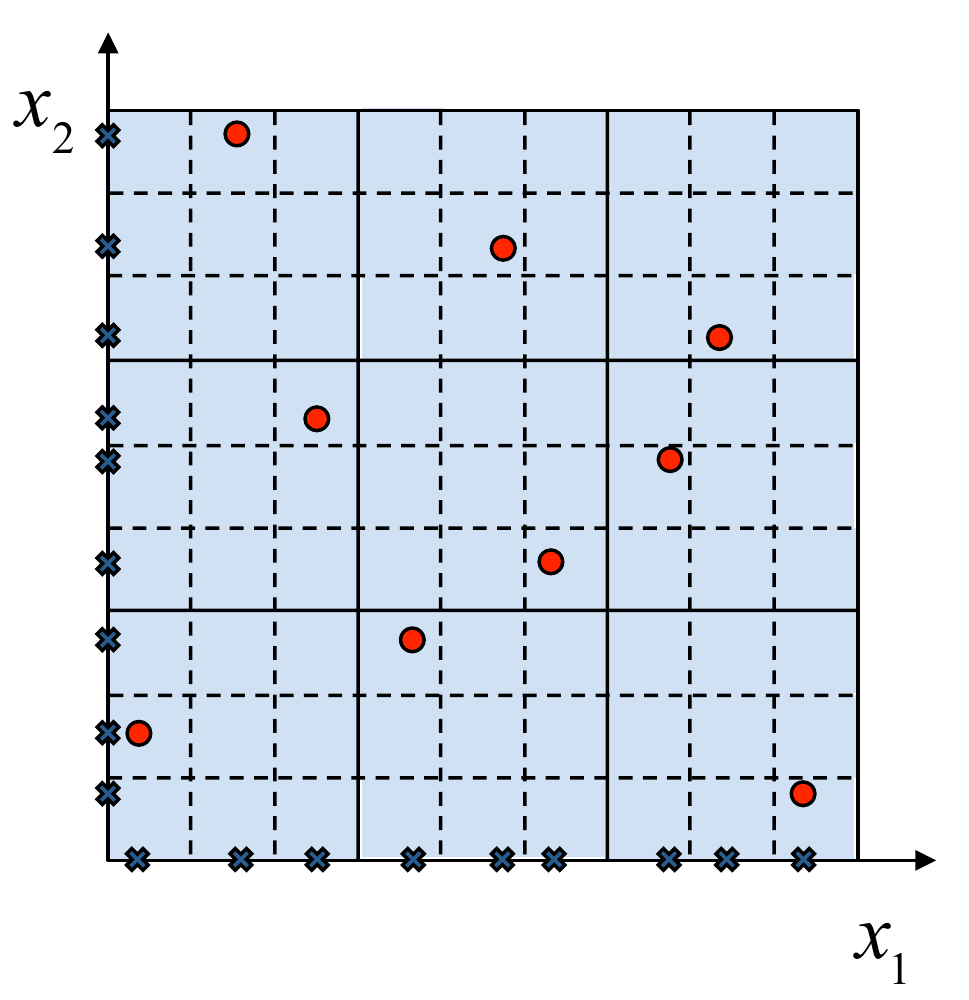}
\caption{Schematic description of the Latinized stratified sampling (``Su-Do-Ku Sampling") method for 9 samples in 2-dimensions.}
\label{fig:LSS}
\end{figure}

It is possible to simultaneously reduce variance associated with the main effects and interaction effects by constructing, on a given $N_i$-dimensional subspace, a true SS that is at the same time a LHS. This is a achieved through a simple procedure referred to herein as ``Latinized" Stratified Sampling (LSS) (colloquially called ``Su-Do-Ku Sampling" as a low dimensional LSS grid resembles the popular number game - see Figure \ref{fig:LSS}).

The LSS method proceeds as follows and is shown graphically in Figure \ref{fig:LSS}:
\begin{enumerate}
\item{Draw a LHS from the subspace - do not group the individual variables (these are shown by an `x' on the ordinates in Figure \ref{fig:LSS}).}
\item Stratify the domain as desired ensuring that the stratification is consistent with an LHS design (solid lines in Figure \ref{fig:LSS}).
\item For each stratum, randomly select a point $x_i$ from each component of the LHS (without replacement) such that the sample $\mathbf{x}=\{x_1,x_2,\dots,x_N\}$ lies within the stratum.
\item Repeat for each stratum of the design. Notice that, after each sample grouping (i.e.\ after a sample is drawn from a given stratum), a number of LHS cells are now ``off-limits" for further sampling. These cells are shaded in Figure \ref{fig:LSS}.
\end{enumerate}

\subsection{Relation to Orthogonal Array-based LHS}
The LSS methodology is not an entirely new development. In fact, an LSS is exactly equivalent to an Orthogonal Array-based LHS (OA-LHS) \cite{Owen_SS_92,Tang_JASA_93} when the strata all have equal probability. A subtle argument is that the LSS approach is perhaps more intuitive than the OA-LHS approach. Certainly, an LSS is easier to construct as it does not depend on generation of large orthogonal arrays that can pose significant challenges. Under the conditions of equivalence between LSS and OA-LHS, the LSS inherits the variance reduction properties that have been theoretically formulated by Tang \cite{Tang_JASA_93}. That is, decomposing the transformation $Y=h(\mathbf{X})$ according to its main effects ($h_i(X_i)=E[h(\mathbf{X}|X_i]-\mu$) and bivariate interactions ($h_{ij}(X_i,X_j)=E[h(\mathbf{X}|X_i,X-j]-\mu-h_i(X_i)-h_j(X_j)$) as:
\begin{equation}
Y = \sum_{i=1}^N h_i(X_i)+\sum_{i=1}^N\sum_{j=1}^{i-1} h_{ij}(X_i,X_j)+r(\mathbf{X}),
\end{equation}
the variance of an OA-LHS/LSS is given by:
\begin{equation}
\text{Var}\{T_O\}=\text{Var}\{T_R\}-\dfrac{1}{n}\sum_{j=1}^N\text{Var}\{f_j(X_j)\}-\dfrac{1}{n}\sum_{i<j}^N\text{Var}\{f_{ij}(X_i,X_j)\}+O(n^{-1})
\end{equation}

The novelty of the proposed approach is that the LSS can be applied to subspaces of different dimension in a simple and efficient way. Therefore, it is capable of targeting specific interactions and reducing variance associated with interactions of different order for different variable combinations without the need to generate large orthogonal arrays. For example, it is capable of reducing variance associated with main effects and interactions among, say, three variables in one subspace while simultaneously reducing main effects and interactions among two variables in another (disjoint) subspace.

\subsection{Challenges revisited}
The use of Latinized Stratified Sampling in combination with Partially Stratified Sampling (LPSS) mitigates the challenges posed in Section 4.3 to a large degree. The primary challenge with PSS is deciding upon a set of appropriate subspaces on which to sample. Using PSS in its basic form, the decision to stratify two variables together based on an assumption that they interact will result in a significant variance cost if those two variables do not actually interact. Meanwhile, the opposite choice will have a similar cost if they do interact. With the introduction of the LPSS, it is sufficient to stratify a set of variables together simply based on the possibility that they may interact. If they do, in fact, interact then the savings will be amplified by reduction in both the main effects and the interactions. If they do not interact, there will be no increase in variance since the main effects are also being filtered.


\section{Demonstration Problems}
In this section, we present a series of high dimensional problems with significant interactions in order to demonstrate the proposed methodology and its potential benefits and shortcomings.

\subsection{High dimensional polynomial functions}
Consider a general high dimensional second order polynomial of the form:
\begin{equation}
\label{eqn:polynomial2}
Y=\sum_{k=1}^{K_{N2}}\alpha_k X_k^2+\sum_{k=1}^{K_I}\beta_k X_{2k-1}X_{2k}+\sum_{k=1}^{K_{N1}}\gamma_k X_k
\end{equation}
where $K_{N2}$ refers to the number of non-interacting second-order terms, $K_I$ refers to the number of second-order interaction terms, and $K_{N1}$ refers to the number of non-interacting first-order terms. The dimension of the polynomial expressed in Eq.\ \eqref{eqn:polynomial2} is equal to $K=\max(K_{N2},2K_I,K_{N1})$. In the following we consider problems of this form with dimension $K=100$ and two different distributions: (a.) $X_k\sim N(0,1)$; and (b.) $X_k\sim N(1,1)$. The shifted mean has the effect of introducing main effects into the second (interaction) term of Eq.\ \eqref{eqn:polynomial2}.

Several cases of the polynomial model given in Eq.\ \eqref{eqn:polynomial2} are studied as outlined in Table 1. In each case, we consider $\alpha_k=\beta_k=\gamma_k=1,\hspace{3pt}\forall k$. For practical purposes the relative influence of each term, as specified by these coefficients, will play an important role in the response $Y$, and by extension will influence the decision to use PSS versus LHS and how to define the PSS subspaces. However, in the interest of conciseness and considering the short study presented in Section 4.2, we focus on the influence of the existence of the various terms and not on their relative strengths. 
\begin{table}
\centering
\label{tab:1}
\begin{tabular} {lccc}
Case & $K_{N2}$ & $K_I$ & $K_{N1}$ \\\hline\hline
1 & 100 & 0, 1, 2, 5, 10, 25, 50 &100\\
2 & 100 & 0, 1, 2, 5, 10, 25, 50  & 0\\
3 & 0 & 0, 1, 2, 5, 10, 25, 50 & 100\\
4 & 0 & 0, 1, 2, 5, 10, 25, 50 & 0\\
\hline
\hline
\end{tabular}
\caption{Dimension and interaction terms for the high dimensional polynomial test problem given by Eq.\ \eqref{eqn:polynomial2}.}
\end{table}

Six different sampling methods have been employed for each case identified in Table 1. In each case 625 samples are drawn. As a baseline for comparison, both SRS and LHS are employed. Two different variations on PSS and LPSS are also used. The first utilizes 2D subspaces for all partial stratification (denoted PSS-$2^{50}$ and LPSS-$2^{50}$ respectively). The second utilizes 2D subspaces for interacting variables and 1D subspaces for non-interacting variables (denoted PSS-$2^{K_I}1^{K-2K_I}$ and LPSS-$2^{K_I}1^{K-2K_I}$). 

\begin{figure}[!ht]
\centering
\subfigure[Case 1\label{fig:N01case1}]{
\centering
\includegraphics[width=0.47\columnwidth]{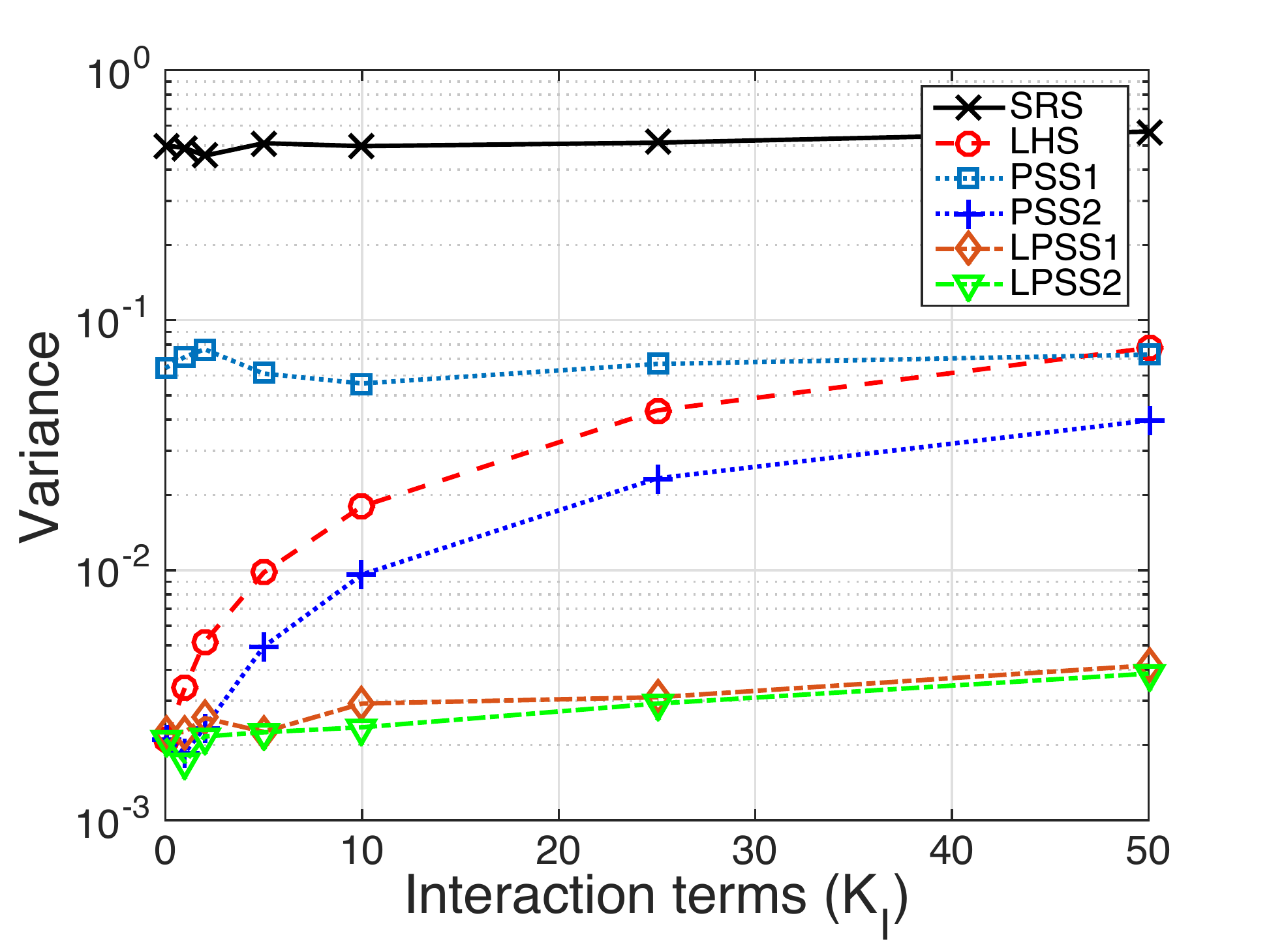}
}
\subfigure[Case 2\label{fig:N01case2}]{
\centering
\includegraphics[width=0.47\columnwidth]{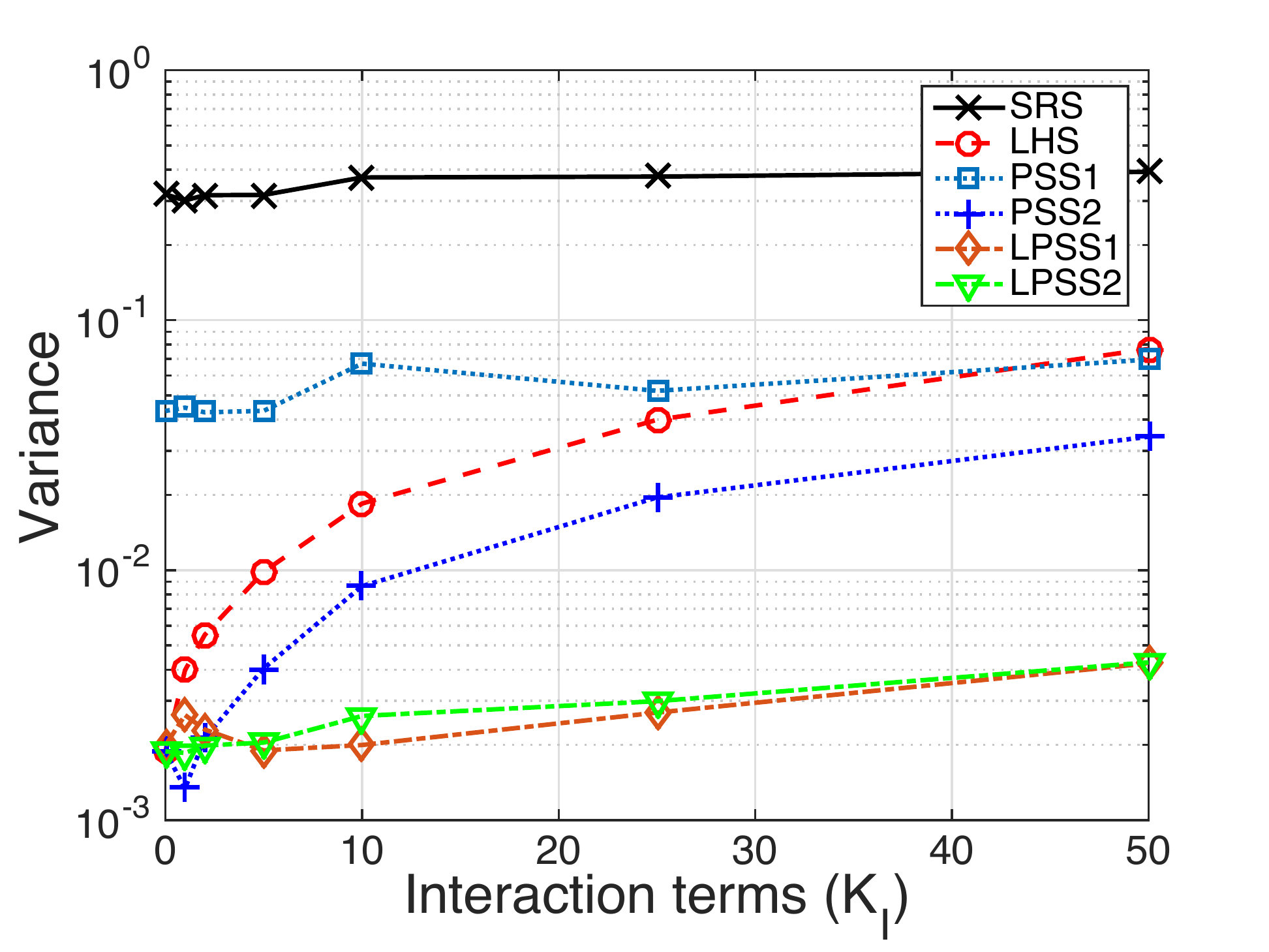}
}
\subfigure[Case 3\label{fig:N01case3}]{
\centering
\includegraphics[width=0.47\columnwidth]{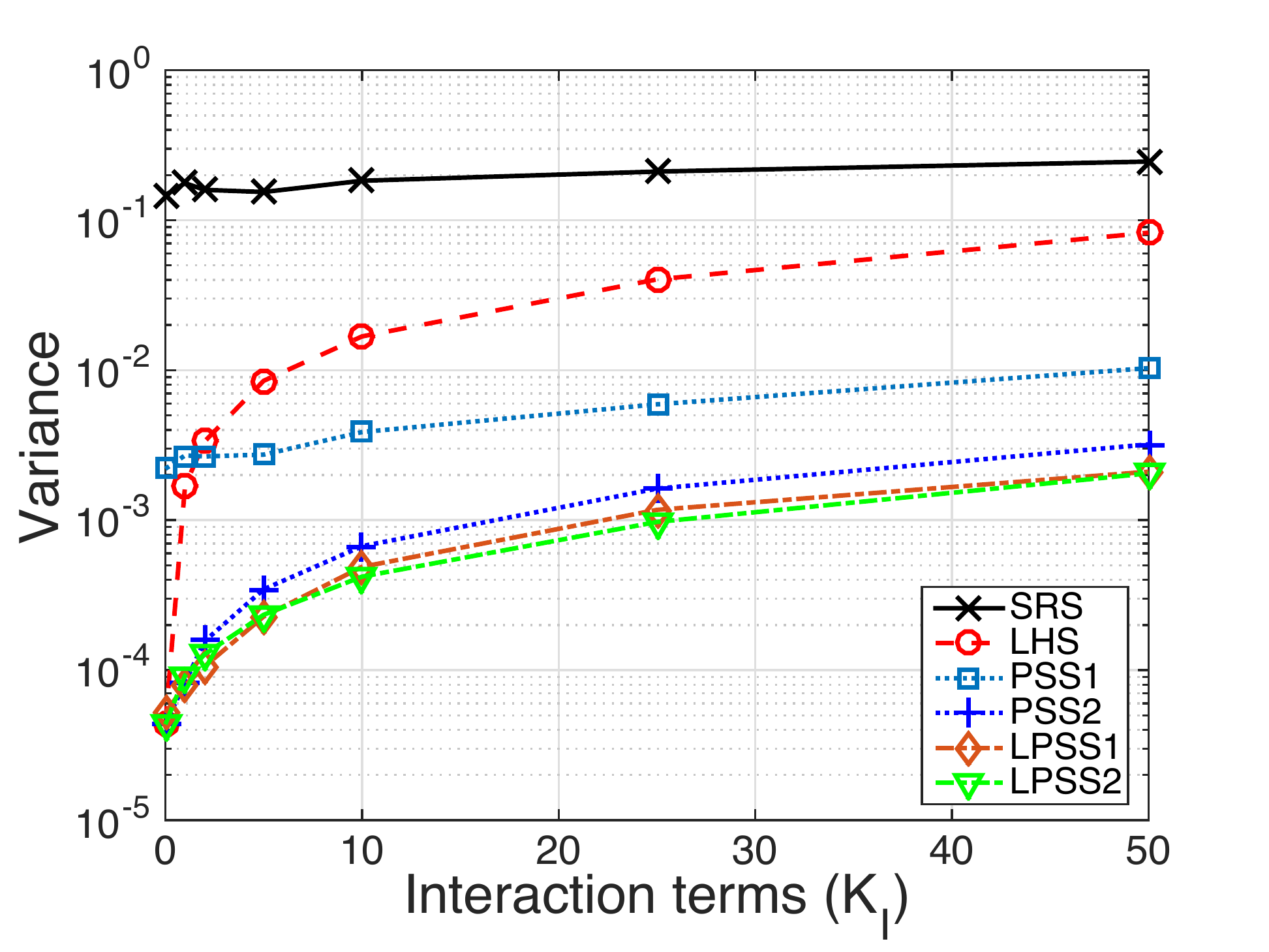}
}
\subfigure[Case 4\label{fig:N01case4}]{
\centering
\includegraphics[width=0.47\columnwidth]{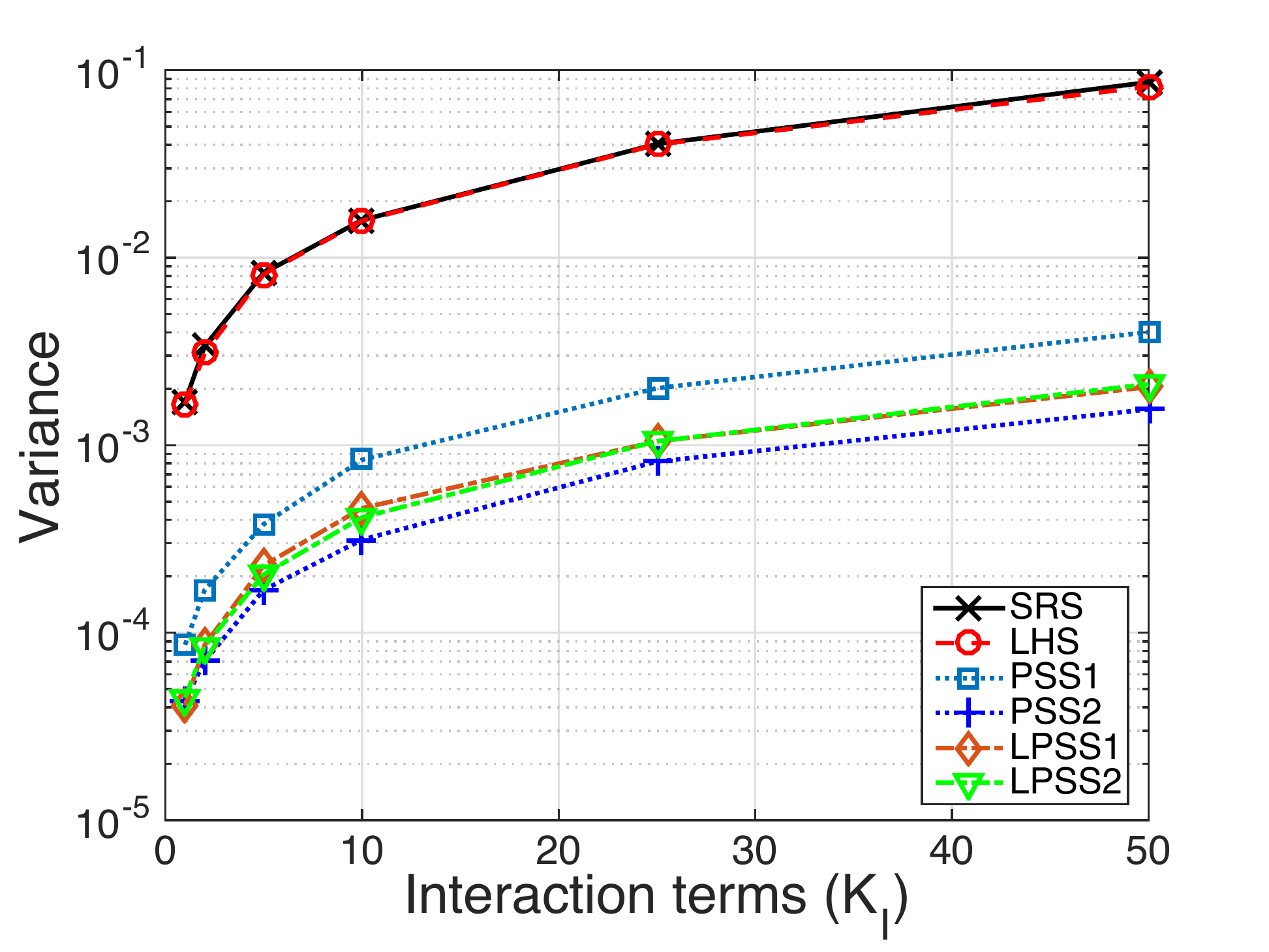}
}
\caption{Variance of Monte Carlo estimates of the mean value for a second-order polynomial function given in Eq.\ \eqref{eqn:polynomial2}. Input variables are N(0,1). Description of the cases are given in Table \ref{tab:1}.}
\label{fig:poly2_N01}
\end{figure}

\begin{figure}[!ht]
\centering
\subfigure[Case 1\label{fig:N11case1}]{
\centering
\includegraphics[width=0.47\columnwidth]{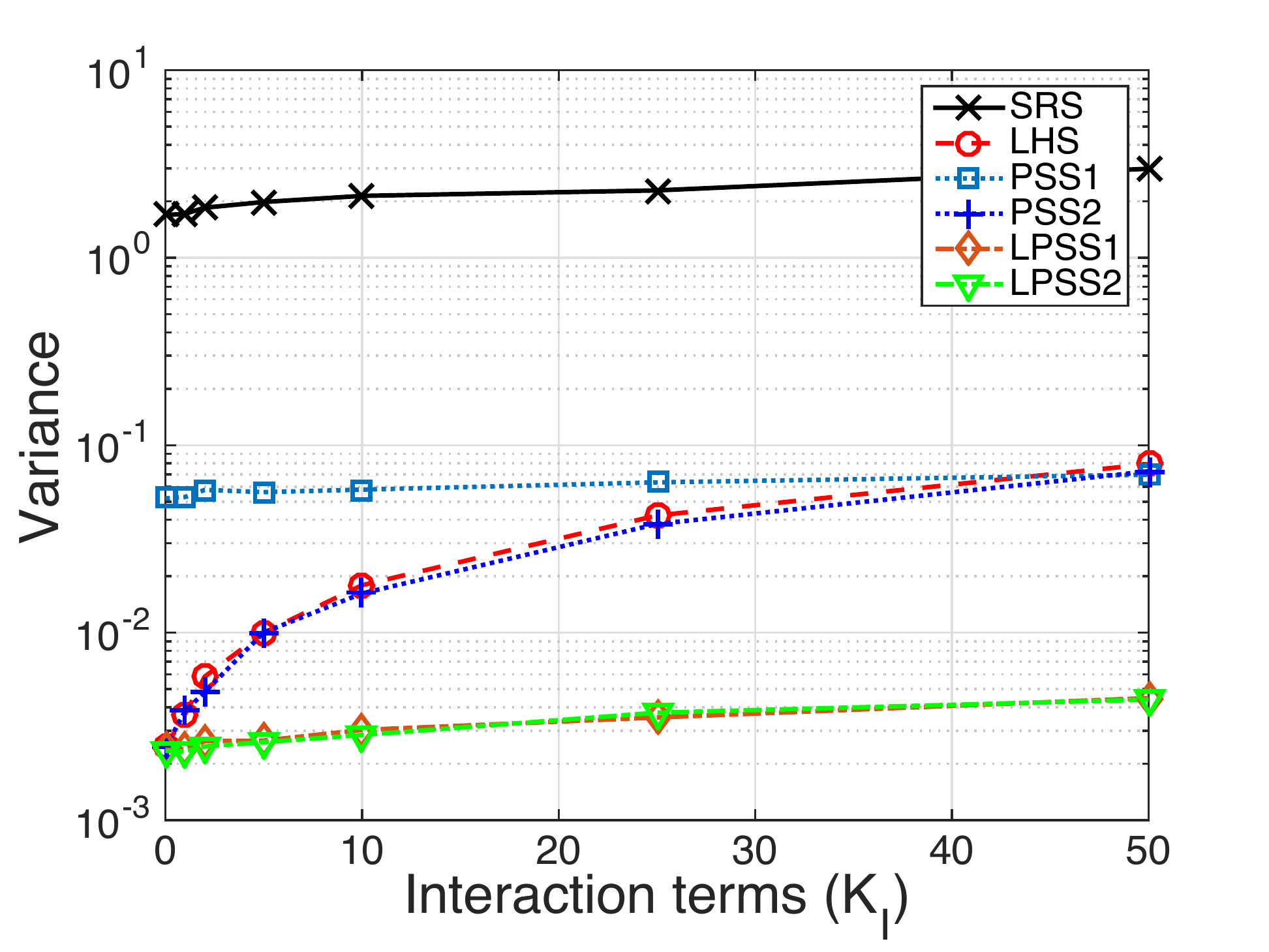}
}
\subfigure[Case 2\label{fig:N11case2}]{
\centering
\includegraphics[width=0.47\columnwidth]{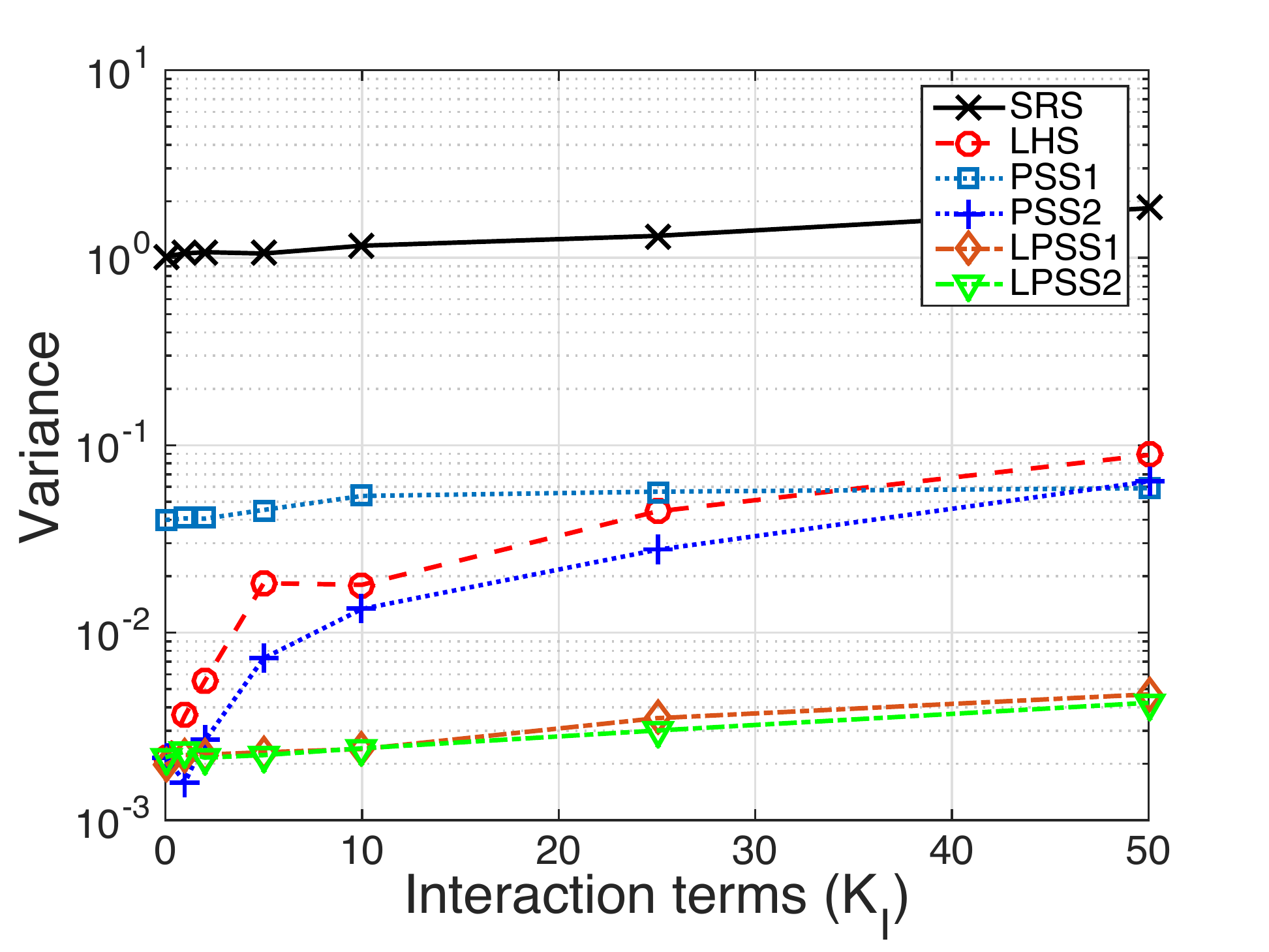}
}
\subfigure[Case 3\label{fig:N11case3}]{
\centering
\includegraphics[width=0.47\columnwidth]{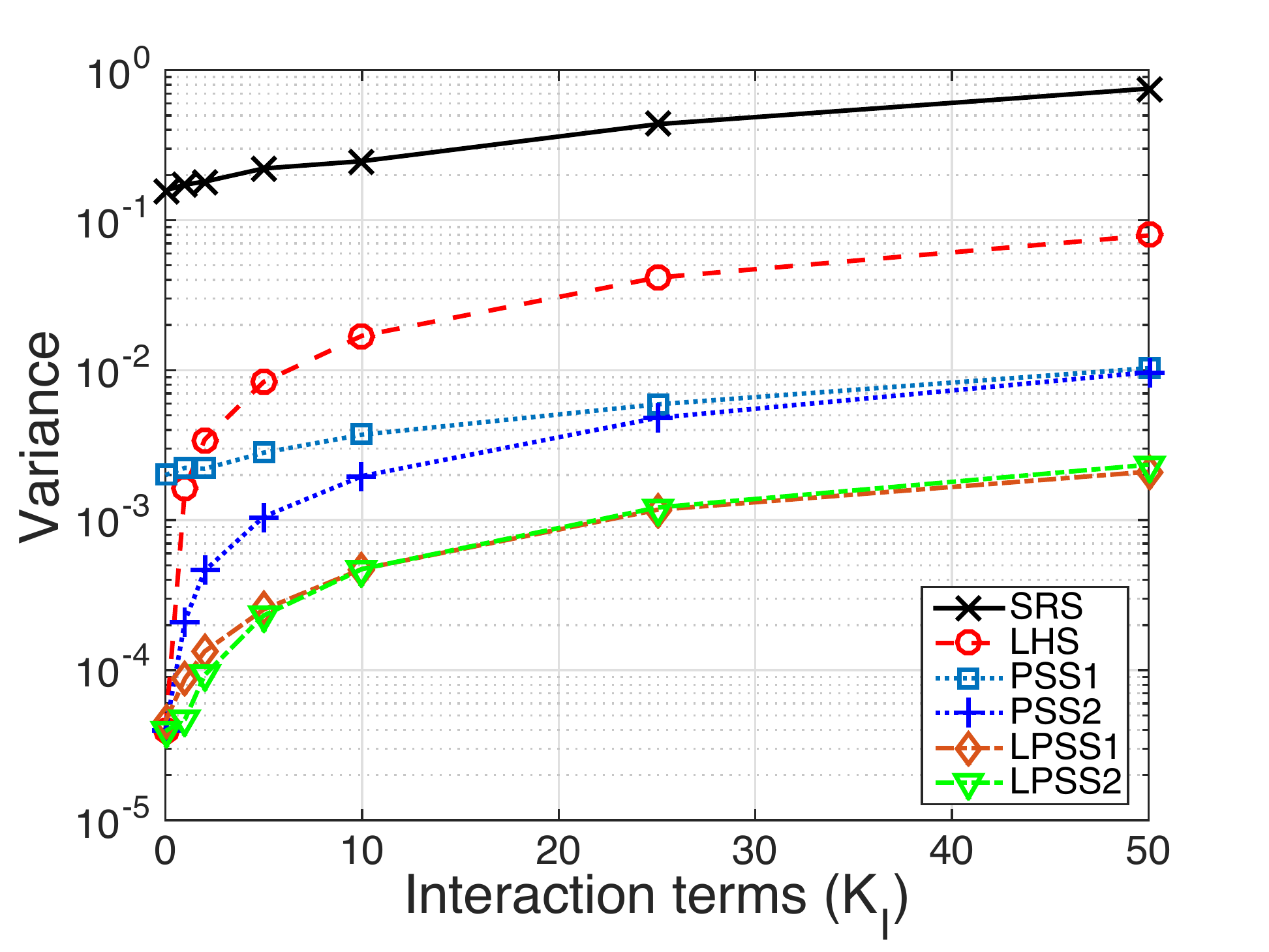}
}
\subfigure[Case 4\label{fig:N11case4}]{
\centering
\includegraphics[width=0.47\columnwidth]{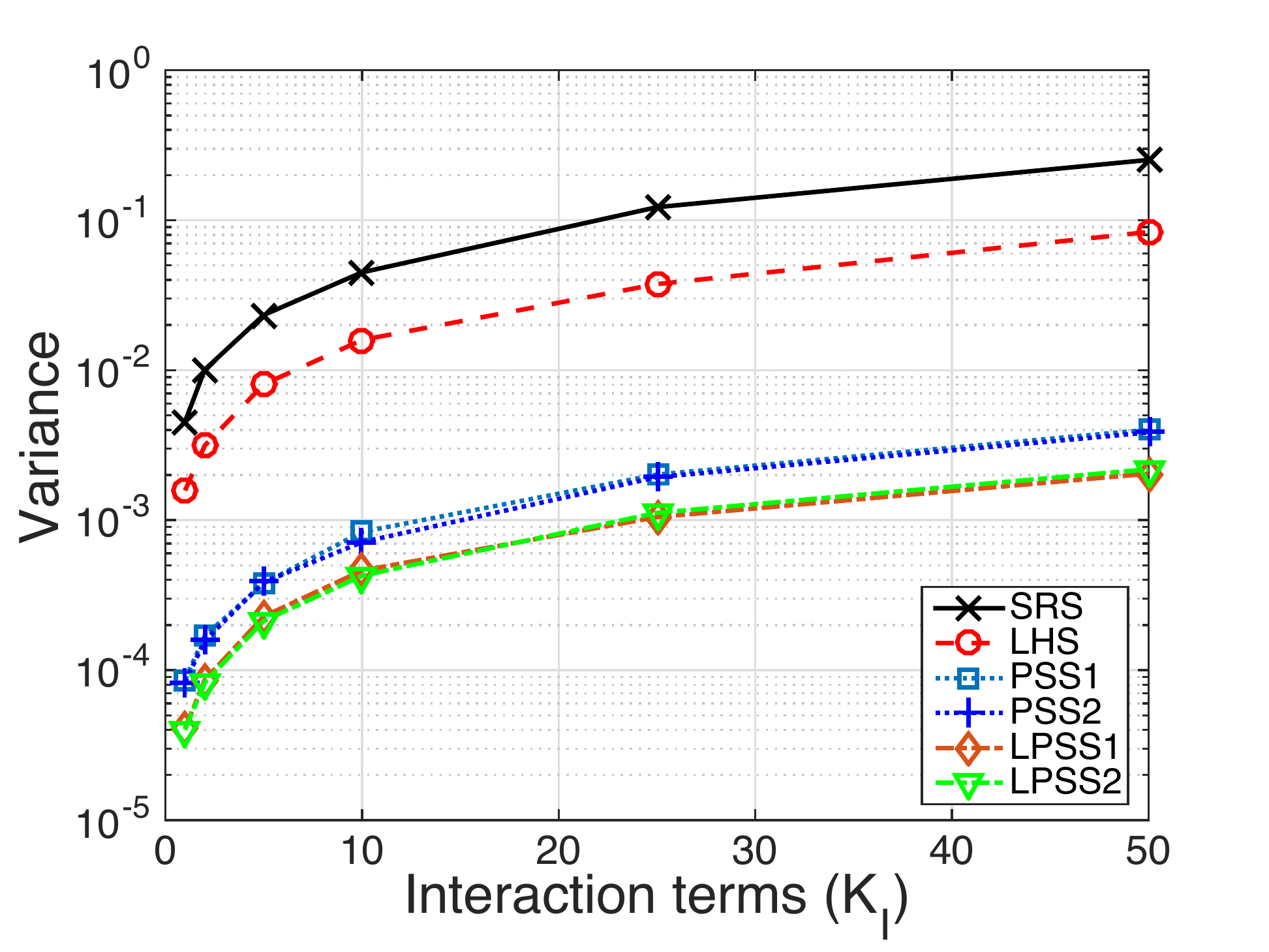}
}
\caption{Variance of Monte Carlo estimates of the mean value for a second-order polynomial function given in Eq.\ \eqref{eqn:polynomial2}. Input variables are N(1,1). Description of the cases are given in Table \ref{tab:1}.}
\label{fig:poly2_N11}
\end{figure}

The variance of the $E[Y]$ estimate from the 625 has been computed for each sampling method from 1000 Monte Carlo simulations. Figure \ref{fig:poly2_N01} plots these variance estimates for cases 1-4 as a function of the number of interaction terms $K_I$ given $X_k\sim N(0,1)$. The same plots for $X_k\sim N(1,1)$ are provided in Figure \ref{fig:poly2_N11}. From these plots, several interesting features are observed.
\begin{itemize}
\item LPSS-$2^{K_I}1^{K-2K_I}$ and LPSS-$2^{50}$ are consistently the best methods in terms of variance reduction. Both perform nearly identically and provide a very large variance reduction - especially when there are both strong main effects and interactions.
\item The variance reduction from PSS-$2^{50}$ depends strongly on the relative strength of the interactions and the main effects. When the main effects are significant PSS-$2^{50}$ is often less effective than LHS. The reason for this is that it stratifies the space based on interactions that are not present at the expense of the main effects in those variables. However, as the interaction strength grows ($K_I$ gets larger), PSS-$2^{50}$ begins to reduce variance compared to LHS. The point at which this occurs depends on the strength of the main effects. For example, in Cases 3 and 4, where the main effects are weak, PSS-$2^{50}$ is nearly always superior to LHS. On the other hand, in Cases 1 and 2 where the main effects are strong, PSS-$2^{50}$ only becomes advantageous for a large number of interaction terms ($K_I=40-50$).
\item As expected from Stein \cite{Stein_Tech_87}, when no main effects are present, LHS does not reduce variance (N(0,1) Case 4 - Fig.\ \ref{fig:N01case4}). However, all PSS and LPSS methods provide considerable variance reduction in this case. 
\end{itemize}

\subsection{Rosenbrock function}

The second problem we consider is the Rosenbrock function \cite{Rosenbrock_TCJ_60} defined as follows:
\begin{equation}
F_R(\mathbf{x})=\sum_{i=1}^{K-1}\left[100(x_i^2-x_{i+1})^2-(x_i-1)^2\right]
\end{equation}
where $\mathbf{x}=\{x_1,x_2,\dots,x_K\}$ is a $K$-dimensional random vector. The Rosenbrock function is a strongly nonlinear high-dimensional function that is commonly used as a benchmark problem for high-dimensional optimization. We consider the case where $K=100$ and $X_i\sim U(0,1)$.

Noticing that, unlike the polynomial function defined in the previous section, although it has only bivariate interactions, these interactions occur between all subsequent variables (e.g. term 1 interacts with term 2 while term 2 independently interactions with term 3). Thus, the choice of partial stratification is less obvious. We consider two different PSS/LPSS designs that group two terms (PSS$-2^{50}$) and four terms (PSS$-4^{25}$) respectively. The mean value $E[F_R]=2013$ is estimated from 625 samples and the standard deviation is computed by repeating the estimation of $E[F_R]$ 5000 times. The results using four different sampling methods (SRS, LHS, PSS, LPSS) are given in Table 2.

\begin{table}
\centering
\label{tab:2}
\begin{tabular} {lccc}
\hline
 & \multicolumn{3}{c}{Standard deviation of mean estimate} \\ \cline{2-4}
Method & Rosenbrock & Schwefel$-N(0,1)$ & Schwefel$-N(1,1)$ \\
 & $E[F_R]=2013$ & $E[F_S]=5057$ & $E[F_S]=343,400$ \\\hline\hline
SRS & 8.778 & 247.5 & 3131.0\\
LHS & 6.756 & 234.1 & 245.9\\
PSS$-2^{50}$ & 4.856 & 227.1 & 358.5\\
PSS$-4^{25}$ & 4.588 & 234.2 & 983.7\\
LPSS$-2^{50}$ & 4.819 & 220.1 & 241.0\\
LPSS$-4^{25}$ & 3.813 & 226.9 & 236.2\\
\hline
\hline
\end{tabular}
\caption{Standard deviation of mean value estimates for the Rosenbrock and Schwefel's functions.}
\end{table}

Notice that, while LHS provides some variance reduction, the PSS and LPSS methods are considerably more effective. Notice also that the 4D partial stratification reduces variance more than the 2D partial stratification. This is because the higher degree of stratification is able to reduce variance associated with more interactions. Finally, similar studies conducted over different ranges of the Rosenbrock function ($X_i\sim U(1,2)$, $X_i\sim U(0,2)$, $X_i\sim U(0,3)$, $X_i\sim U(0,5)$, $X_i\sim U(9,10)$) yield trends consistent with those in Table 2.

\subsection{Schwefel's Problem 1.2}

The final numerical demonstration problem is the Schwefel's problem 1.2 that possesses very high dimensional interactions defined by:
\begin{equation}
F_S(\mathbf{x}) = \sum_{i=1}^K\left(\sum_{j=1}^ix_j\right)^2
\label{eq:schwefel}
\end{equation}
Again, we consider $K=100$ and consider the standard deviation of estimates of the expected value from 5000 repeats of 625 samples. As summarized in Table 2, we consider $X_i\sim N(0,1)$ and $X_i\sim N(1,1)$ with true mean values estimated from Monte Carlo simulation of $5,057$ and $343,400$ respectively and two different partial stratifications (PSS/LPSS$-2^{50}$ and PSS/LPSS$-4^{25}$). In the case of $X_i\sim N(0,1)$, it is clear from Eq.\ \eqref{eq:schwefel} that the main effects are very small. Hence, not only is the standard deviation very large (yielding a coefficient of variation on the estimate of nearly 5\%), but LHS produces essentially negligible variance reduction. Similarly, the PSS and LPSS methods are equally ineffective at reducing variance in this problem. However, given a shifted mean $X_i\sim N(1,1)$, expansion of Eq.\ \eqref{eq:schwefel} yields significant main effects. Thus, LHS produces more than an order of magnitude variance reduction. PSS is slightly less effective since it is able to reduce variance associated with the $\le$2nd and $\le$4th order interactions respectively but cannot reduce variance associated with the significant main effects. The LPSS method remains effective because it reduces variance associated with both effects. 

This problem serves to illuminate the limitations of both the proposed methods and existing variance reduction techniques. There is currently no effective means of reducing variance associated with very high-dimensional interactions. Fortunately, such high-dimensional interactions typically occur only in select circumstances.

\section{Application to plate buckling}
The assessment of structural component strength often involves the interaction of various material and geometry parameters. One such case is the analysis of plate buckling strength which is of interest in many structural applications. Consider a rectangular plate that is simply supported on all four edges subjected to uniaxial compression. Analytical expressions have been derived to assess the buckling strength, first by Faulkner \cite{Faulkner_JSR_75}, who derived the normalized buckling strength for a pristine plate as:
\begin{equation}
\phi=\dfrac{\sigma_u}{\sigma_0}=\left(\dfrac{2}{\lambda}-\dfrac{1}{\lambda^2}\right)
\end{equation}
where $\sigma_u$ is the ultimate load capacity, $\sigma_0$ is the yield stress of the material, and
\begin{equation}
\lambda=\dfrac{b}{t}\sqrt{\dfrac{\sigma_0}{E}}
\label{eq:slenderness}
\end{equation}
is referred to as the slenderness of the plate with width $b$, thickness $t$, and elastic modulus $E$. This equation was later modified by Carlsen \cite{Carlsen_NMR_77} to include the effects of non-dimensional initial deflections $\delta_0=\frac{w_0}{t}$ where $w_0$ is the magnitude of the deflection and residual stresses resulting from welding at the perimeter as:
\begin{equation}
\phi=\left(\dfrac{2.1}{\lambda}-\dfrac{0.9}{\lambda^2}\right)\left(1-\dfrac{0.75\delta_0}{\lambda}\right)\left(1-\dfrac{2\eta t}{b}\right)
\label{eqn:buckling}
\end{equation}
where $\eta t$ is the width of the zone of tension residual stresses. 

Uncertainty analysis of such plates has long been of interest to the naval community \cite{Guedes_Soares_SS_88} and it is clear from Eq.\ \eqref{eqn:buckling} that the various parameters controlling the buckling strength and its variability interact strongly. We are interested in defining the ``best" stratification of the domain such that variance of response statistics from a Monte Carlo analysis of plate buckling will be minimized. We consider mild structural steel plates commonly used in naval applications with material and geometric variabilities estimated from the data presented by Hess et al. \cite{Hess_et_al_NEJ_02} and Guedes Soares \cite{Guedes_Soares_SS_88} as presented in Table 3. The plate width $b$ is estimated based on statistical analysis of variability in stiffener spacing for common ship plates. Thickness variations are estimated from statistical analysis of plate thickness from over 2000 measurements \cite{Hess_et_al_NEJ_02} while elastic modulus and yield stress result from statistical analysis of common structural steel such that the large mean value and COV for yield stress is a result of the common practice of downgrading higher strength steels that do not meet specification and the fact that Navy standards require only a minimum yield stress. Lastly, the initial deflections and residual stress variations are prescribed according to the findings of Faulkner \cite{Faulkner_JSR_75} and Antonia \cite{Antoniou_JSR_80} as reported by Guedes Soares \cite{Guedes_Soares_SS_88}.
\begin{table}
\centering
\resizebox{\columnwidth}{!}{
\label{tab:3}
\begin{tabular} {cccccc}
\hline
Variable & Physical Meaning & Nominal Value & Mean & COV & Distribution Type \\\hline\hline
$b^*$ & width (in.) & 24 & $0.992\times 24$ & 0.028 & Normal\\
$t^*$ & thickness (in.) & 0.5 & $1.05\times 0.5$ & 0.044 & LogNormal\\
$\sigma_0^*$ & yield stress (ksi) & 34 & $1.3\times 34$ & 0.1235 & LogNormal\\
$E^*$ & elastic modulus (ksi) & 29,000 & $0.987\times 29,000$ & 0.076 & Normal\\
$\delta_0=\dfrac{w_0}{t}$ & initial deflection & 0.35 & $1.0 \times 0.35$ & 0.05 & Normal\\
$\eta$ & residual stress & 5.25 & $1.0\times 5.25$ & 0.07 & Normal\\
 \hline
\hline
\end{tabular}
}
\caption{Distributions for plate material, geometry, and imperfection variables. Distributions based on guidance from \cite{Hess_et_al_NEJ_02} and \cite{Guedes_Soares_SS_88}.}
\end{table}

As illustrated in Table 3, the plate buckling problem considered here is 6-dimensional. Thus, there are many possible partial stratifications. We consider three different PSS/LPSS designs (4 if LHS is included) as shown in Table 4 that attempt to group variables according to the interactions suggested by Eqs.\ \eqref{eq:slenderness} and \eqref{eqn:buckling}. In each case, a total of 625 samples are drawn. For this problem, the variance reduction associated with filtering the main effects is more significant than that of filtering just the interactions. That is, LHS provides a greater variance reduction than all three PSS designs. However, the problem is able to benefit from reducing the variance associated with both main effects and interactions as all three LPSS designs provide a greater variance reduction than LHS, with LPSS$-4^11^2$ providing the largest variance reduction. Thus, we conclude that the LPSS-$-4^11^2$ is the most effective sample design for this problem. Referring to the discussion of Section 3.3, this design has a similar effect (though not exactly) to constructing a Latin hypercube sample on the three variables $\lambda$, $\delta_0$, and $\eta$. Given, the form of Eq.\ \eqref{eqn:buckling}, this is somewhat intuitive.

\begin{table}
\centering
\resizebox{\columnwidth}{!}{
\label{tab:4}
\begin{tabular} {lcccc}
\hline
Design & Paired Variables & Strata & Mean Strength $E[\phi]$ & Std. Dev. of $E[\phi]$ \\\hline\hline
SRS & N/A & N/A & 0.5590 & $2.00e-3$\\
LHS & $[b]; [t]; [\sigma_0]; [E]; [\delta_0]; [\eta]$ & $[625,625,625,625,625,625]$& 0.5590 & $4.33e-4$ \\
PSS$-2^{3}$ & $[b,t]; [\sigma_0,E]; [\delta_0,\eta]$ & $[25,25,25]$ & 0.5590 & $6.36e-4$  \\
PSS$-2^{2}1^2$ & $[b,t]; [\sigma_0,E]; [\delta_0]; [\eta]$ & $[25,25,625,625]$ & 0.5590 & $6.17e-4$ \\
PSS$-4^{1}1^2$ & $[b,t,\sigma_0,E]; [\delta_0]; [\eta]$ & $[5,625,625,625]$ & 0.5590 & $1.10e-3$ \\
LPSS$-2^{3}$  & $[b,t]; [\sigma_0,E]; [\delta_0,\eta]$ & $[25,25,25]$ & 0.5590 & $3.87e-4$ \\
LPSS$-2^{2}1^2$ & $[b,t]; [\sigma_0,E]; [\delta_0]; [\eta]$ & $[25,25,625,625]$ & 0.5590 & $3.95e-4$ \\
LPSS$-4^{1}1^2$ & $[b,t,\sigma_0,E]; [\delta_0]; [\eta]$ & $[5,625,625,625]$ & 0.5590 & $3.45e-4$ \\
 \hline
\hline
\end{tabular}
}
\caption{Distributions for plate material, geometry, and imperfection variables. Distributions based on guidance from \cite{Hess_et_al_NEJ_02} and \cite{Guedes_Soares_SS_88}.}
\end{table}

\section{Conclusions}

In this work, the Latin hypercube sampling method of \cite{McKay_et_al_Tech_79} has been generalized by defining a spectrum of stratified sampling methods of which true stratified sampling and Latin hypercube sampling lie at its extremes. The intermediate designs on the spectrum are defined as Partially Stratified Sample designs and are shown to reduce variance associated with variable interactions whereas LHS reduces the variance associated with the main effects of the variables. The properties of PSS designs are studied and some considerations and challenges associated with their use are discussed. The challenges and shortcomings of PSS designs are mitigated by coupling a PSS design with a newly introduced stratified sampling method called Latinized stratified sampling that is simultaneously a LHS design and a fully stratified design. The LSS method is equivalent, under certain conditions, to an Orthogonal Array based Latin hypercube but is signficantly simpler to obtain. Merging the LSS method with the PSS method yields the Latinized Partially Stratified Samping method (LPSS) which is shown, for many high dimensional applications, to provide superior variance reduction when both low-order interactions and main effects are present. Several numerical examples have been provided and the new methods have been applied to a plate buckling problem in structural mechanics. 

\bibliographystyle{elsarticle-num}
\bibliography{bibliography}{}

\end{document}